\documentclass{aa} 
\usepackage{natbib}
\usepackage{graphicx}
\usepackage{txfonts}
\usepackage{epsfig}
\usepackage{times}
\usepackage{rotating}
\usepackage{float}
\usepackage[caption = false]{subfig}
\usepackage{setspace}
\usepackage{lscape}
\usepackage{epstopdf}
\usepackage{tabularx}
\usepackage{caption}
\usepackage[usenames]{color}
\usepackage[toc,page]{appendix}
\usepackage{soul}

\DeclareCaptionFormat{cont}{#1 (cont.)#2#3\par}

\begin{document}

\title{Fossil group origins}

\subtitle{XII. Large-scale environment around fossil systems}

\authorrunning{S. Zarattini et al.}

\titlerunning{Large-scale environment around FGs}

\author{S. Zarattini\inst{1,2}, J. A. L. Aguerri\inst{3,4},  R. Calvi\inst{3,4}, M. Girardi\inst{5,6}}

\institute{Dipartimento di Fisica e Astronomia ``G. Galilei'', Universit\`a di Padova, vicolo dell'Osservatorio 3, I-35122 Padova, Italy \\
\email{stefano.zarattini@unipd.it}
\and INAF - Osservatorio Astronomico di Padova, vicolo dell'Osservatorio 2, I-35122 Padova, Italy
\and Instituto de Astrof\'isica de Canarias, calle Vía L\'actea s/n, E-38205 La Laguna, Tenerife, Spain
\and Departamento de Astrof\'isica, Universidad de La Laguna, Avenida Astrof\'isico Francisco S\'anchez s/n, E-38206 La Laguna, Spain
\and INAF-Osservatorio Astronomico di Trieste, via Tiepolo 11, I-34143 Trieste, Italy
\and Dipartimento di Fisica, Universit\`{a} degli Studi di Trieste, via Tiepolo 11, I-34143 Trieste, Italy}

\date{\today}

\abstract{}
{We analyse the large-scale structure out to 100 Mpc around a sample of 16 confirmed fossil systems using spectroscopic information from the Sloan Digital Sky Survey Data Release 16.}
{We compute the distance between our FGs and the centres of filaments and nodes presented in \citet{Chen2016}. We also study the density of bright galaxies, since they are thought to be good mass tracers, and the projected over densities of galaxies. Finally, we apply a FoF algorithm to detect virialised structures around our FGs, in order to have an estimate of the mass available in their surroundings.}
{FGs are mainly found close to filaments, with a mean distance of $3.7 \pm 1.1$ R$_{200}$ and a minimum distance of 0.05 $R_{200}$. On the other hand, none of our FGs is found close to intersections, with a mean and minimum distance of $19.3 \pm 3.6$ and 6.1 $R_{200}$, respectively.
There is a correlation for which FGs at higher redshifts are found in denser regions, when we use bright galaxies as tracers of the mass. At the same time, FGs with the largest magnitude gaps ($\Delta m_{12}$ > 2.5) are found in less dense environments and hosting, on average, smaller central galaxies.}
{Our results suggest that FGs formed in a peculiar position of the cosmic web, close to filaments and far from nodes, in which their interaction with the cosmic web itself can be limited. We deduce that FGs with faint BCGs, large $\Delta m_{12}$, and low redshifts could be systems at the very last stage of their evolution. Moreover, we confirm theoretical predictions that systems with the largest magnitude gap are not massive.}


\maketitle

\section{Introduction}
\label{sec:intro}
Fossil groups (FGs) were proposed by \citet{Ponman1994}, when they found an apparently isolated giant elliptical galaxy that was surrounded by an X-ray halo, typical of a group of galaxies. Their interpretation of these observations was that this system was the final stage of the evolution of a group of galaxy, in which all the other $M^*$ galaxies (where $M^*$ is the characteristic luminosity of the luminosity function) were merged within the brightest central galaxy (BCG). To accomplish this scenario, FGs were supposed to be older than regular groups and to remain isolated from the cosmic web. In this picture, FGs can be considered fossil relics of the primordial Universe.

Only in the last decade the number of known FGs grew enough so to have systematic studies of these objects. Without wanting to be exhaustive, four over-luminous red galaxies were found from \citet{Vikhlinin1999}, five FGs were presented in \citet{Jones2003}, 34 FG candidates were proposed in \citet{Santos2007} \citep[with 15 confirmed in][]{Zarattini2014}, 12 new FGs were presented in \citet{Miller2012}, 18 FG candidates were presented in \citet{Adami2018} and in \citet{Adami2020} a novel probabilistic approach was used to favour statistical studies of FGs. Recently, a list of 36 confirmed FGs (taken from the literature) was presented in the review of in \citet{Aguerri2021}, spreading on the redshift range $0 \le z \le 0.5$.

The hierarchical model of structure formation in the Universe is a remarkably successful theory. It predicts that small structures form earlier and that they collapse subsequently into larger structures. This model is proven to be successful at large scales, but at galactic scales it incurs in the so-called ``small-scale crisis'': the number of predicted low-mass sub-halos in simulations around Milky-way like galaxies is larger than the one observed. On the other hand, \citet{D'Onghia2004} found that the small-scale crisis could be affecting the larger halos of FGs and that, in this case, the missing satellites can be as massive as the Milky Way itself. However, \citet{Zibetti2009} and \citet{Lieder2013} found no signs of missing satellites in FGs.

The Fossil Group Origins project \citep[FOGO,][]{Aguerri2011} is devoted to the study of one of the largest sample of FGs available in the literature. The previous eleven publications of the FOGO team shed light onto various aspects of FGs. In particular, the formation of their BCGs were studied in \citet{Mendez-Abreu2012}, whereas their stellar populations were analysed in \citet{Corsini2018}. In these works, the authors showed that BCGs in FGs are amongst the most-massive galaxies in the Universe and that their stellar age is compatible with central galaxies in non-fossil systems. At the same time, BCGs are found to be more segregated in the velocity space when compared to non-FGs \citep{Zarattini2019}. The scaling relations between optical and X-ray \citep{Girardi2014,Kundert2015} showed that FGs were not X-ray over luminous systems and that particular attention must be dedicated to the homogeneity of the data in these kind of studies. Also, the luminosity functions (LFs) of FGs were compared with those of non-FGs \citep{Zarattini2015,Aguerri2018}, finding that there is a dependence of the faint-end slope on the magnitude gap. Moreover, FGs were found to host a similar fraction of substructures as non-FGs \citep{Zarattini2016}. Finally, in \citet{Zarattini2021} we showed that one of the main differences between fossil and non-fossil systems can be found in the different orbital shape of their galaxies. In fact, we found that galaxies falling into FGs are found on more radial orbits than in non-FGs, in agreement with theoretical predictions \citep{Sommer-Larsen2005}.

Few works were devoted to the study of the large-scale environment around FGs. In particular, \citet{Adami2012} compared the environment of two FGs with one non-FG using photometric data. They found FGs to be more isolated than the control cluster, but the statistic was quite low and prevented them from reaching general conclusion. In a similar way, \citet{Adami2018} used spectroscopic data of RXJ1119.7+2126 (one of the two FGs of their previous paper) and were able to confirm that this system is located in a poor environment.

The aim of this work is thus to analyse the large-scale structure around a large sample of spectroscopically-confirmed FGs in order to understand if they are found in any peculiar position of the cosmic web.

The cosmology used in this paper, as in the rest of the FOGO publications, is H$_0 = 70$ km s$^{-1}$, $\Omega_\lambda = 0.7$, and $\Omega_M = 0.3$.
\section{Sample selection}
\label{sec:sample}
Our sample is selected from the review of \citet{Aguerri2021}, where a list of spectroscopically-confirmed FGs in presented in table 1. In particular, we chose all the systems which centres are found in the Sloan Digital Sky Survey Data Release 16 (SDSS DR16) footprint and with an upper redshift limit of $z=0.20$. The SDSS spectroscopy is mainly limited at $m_r = 17.77$, that is equivalent to $M_r \sim -22$ at $z=0.2$. The central galaxies of this sample have magnitude ranging between $M_r = -21.3$ and $M_r = -24.1$, the faintest of the sample would not be mapped at $z=0.2$. We thus decided to limit the sample to $z=0.15$, corresponding to a magnitude limit of $M_r \sim -21.5$. The systems selected according to these criteria are 18. We note that we also included SDSS J1045+0420, which redshift is $z=0.154$.

\begin{figure}
    \centering
    \includegraphics[trim=60 350 50 80,width=0.5\textwidth]{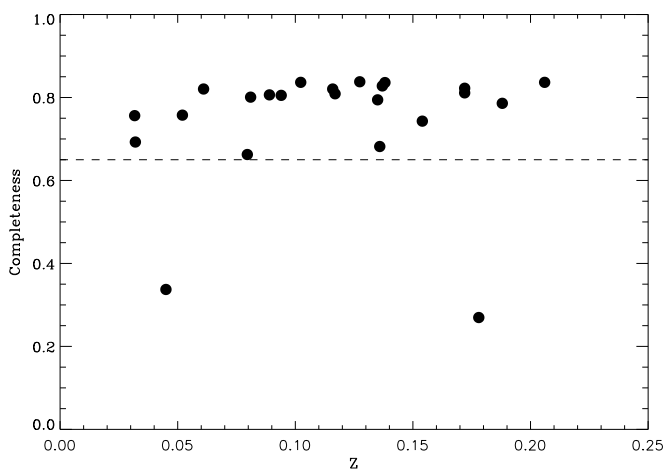}
    \caption{The spectroscopic completeness for our starting sample. The two systems with completeness smaller than 65\% were not included in our final sample.}
    \label{fig:completitud}
\end{figure}

We thus looked at the spectroscopic completeness of the various clusters. In fact, it is possible that some cluster is part of the SDSS footprint, but for some reason its spectroscopy is below the standard. We computed the fraction of galaxies with spectroscopy by considering the number of galaxies with spectroscopy and the number of targets (e.g. galaxies with $m_r \le 17.77$). In Fig. \ref{fig:completitud} only two clusters have completeness lower than 65\%: these are UGC 842 and 1RXS J235814.4 + 150524. The former is found in a region of SDSS with very uneven spectroscopic coverage, the latter would also be discarded using the redshift criterium, so we did not check its spectroscopic coverage in details. We remain with 17 FGs after applying this cut.

Finally, we will explain in Sect. \ref{sec:FGS28} that we prefer to exclude FGS28 from our sample, due to its doubtful nature.

Our final sample is thus composed of 16 FGs which properties are presented in Table \ref{tab:sample}. We note that the coordinates reported in the table are those of the brightest central galaxy (BCG). This was done for having homogeneous centres. In fact, for some systems (like the FOGO ones) the centre reported in the literature is that of the BCG, for some others the centre reported in the literature is obtained from X-ray data. There is a known discrepancy between these two centres, estimated in 13 kpc in \citet{Sanderson2009} using a sample of 65 X-ray-selected clusters and in less than 5\% of $R_{200}$ from \citet{Lin2004}. Moreover, \citet{Aguerri2007} compared the distance between the X-ray peak and the galaxy surface density, finding a mean difference of 150 kpc. We don't expect this difference to impact in the results of our work, since all our tests involve megaparsec scales. Moreover, magnitude gaps (within 0.5 $R_{200}$) and X-ray luminosities are also taken directly from \citet[][]{Aguerri2021} and references therein. However, in Sect. \ref{sec:lx} we will explain how we converted the $L_X$ values to bolometric $L_X$ luminosities and to the same cosmology used in this paper. On the other hand, few masses were available in \citet[][]{Aguerri2021} and references therein, so we decided to estimate them by using their bolometric X-ray luminosities, thus homogenising the computation.

The SDSS catalogues were obtained with an SQL query to the CasJobs webpage \footnote{\url{http://skyserver.sdss.org/CasJobs/default.aspx}}. For each system, we looked for all galaxies with available spectroscopy within a radius of 100 Mpc from the centre of the system reported in Table \ref{tab:sample}

It is worth noting that for some cluster it was not possible to obtain data for the entire 100-Mpc-radius area. In Appendix \ref{appendix} we present the entire field of view available for each cluster. For 8 clusters, the full area is covered (11 if we consider a circle of 50 Mpc radius). For the remaining clusters the coverage is not full, but we can still use them for the majority of our scopes.

\subsection{$L_X$ luminosities}
\label{sec:lx}

We were able to obtain the X-ray luminosity for all the systems in the sample. All the $L_X$ were taken from the literature and were computed in different bands, cosmologies, and methods. We then converted all the $L_X$ to the same band and cosmology. In particular, we chose to convert to bolometric luminosity and to the cosmology of this paper, when needed.

Bolometric luminosities were obtained by multiplying the luminosity in the different bands by a temperature-dependent correction factor. In particular, we found luminosities in the $0.1-2.4$ keV (two FGs) and $0.5-2.0$ keV bands (seven FGs); the remaining seven FGs were found already in bolometric. 
This factor was computed using the Raymond-Smith code with ICM of 0.3 $Z_\odot$, according to the PIMMS\footnote{\url{https://cxc.harvard.edu/toolkit/pimms.jsp}} webpage. 

The code needs the value of the X-ray temperature T$_{\rm X}$. The values of T$_{\rm X}$ are available in the literature for seven out of nine clusters. For the remaining two clusters we followed a recursive procedure based on the T$_{\rm X}$-L$_{\rm X}$ relation \citep[see][]{Girardi2014}.
The values of the (bolometric) X-ray luminosities are listed in Table 1.

\begin{table*}  
\caption{Global properties of the sample}
\label{tab:sample}
\begin{center}
\tiny
\begin{tabular}{lccccccccccc}
\hline 
\noalign{\smallskip}   
%
%
\multicolumn{1}{l}{Cluster name} & \multicolumn{1}{c}{R.A.} & \multicolumn{1}{c}{Dec} & \multicolumn{1}{c}{$\Delta m_{12}$} & \multicolumn{1}{c}{$\Delta m_{14}$} & \multicolumn{1}{c}{M$_{r,BCG}$} & \multicolumn{1}{c}{$z_{\rm \,BCG}$} & \multicolumn{1}{c}{M$_{200}$}  & \multicolumn{1}{c}{R$_{200}$} & \multicolumn{1}{c}{L$_{\rm X}$} & \multicolumn{1}{c}{ L$_{\rm X}$ source} \\
\multicolumn{1}{l}{} & \multicolumn{1}{c}{[Deg]} & \multicolumn{1}{c}{[Deg]} & \multicolumn{1}{c}{} & \multicolumn{1}{c}{} & \multicolumn{1}{c}{[Mag]} & \multicolumn{1}{c}{} & \multicolumn{1}{c}{[$10^{14}$ M$_\odot$]} & \multicolumn{1}{c}{[Mpc]} & \multicolumn{1}{c}{[$10^{42}$ erg s$^{-1}$]} & \multicolumn{1}{c}{}  \\
\multicolumn{1}{l}{(1)} & \multicolumn{1}{c}{(2)} & \multicolumn{1}{c}{(3)} & \multicolumn{1}{c}{(4)} & \multicolumn{1}{c}{(5)} & \multicolumn{1}{c}{(6)} & \multicolumn{1}{c}{(7)} & \multicolumn{1}{c}{(8)} & \multicolumn{1}{c}{(9)} & \multicolumn{1}{c}{(10)} & \multicolumn{1}{c}{(11)}\\
\noalign{\smallskip}  
\hline
DMM2008 IV & 113.641000 & 26.899000	 & 2.4	 & 3.0 & -23.43 & 0.08	 & 0.34 & 0.67	 & 41.8 & [1]\\
FGS03 & 118.184151 & 45.94928 & 2.09 & 2.55	 & -22.40 & 0.05 & 1.01	 & 0.96	 & 29.3 &  [2]\\
SDSS J0906+0301 & 136.659489 &	3.0275479	& 3.09 & / & -23.47 & 0.14	& 0.33	& 0.66	& 4.8 &  [3] \\
A1068 & 160.182917	 & 39.948056 & 2.3	 & 3.1 & -23.99 & 0.14	 & 2.05 & 1.22 & 2388.5	&  [1]\\
SDSS J1045+0420 & 161.452085 	 & 4.3423831 & 2.00 & /	 & 	-23.54 & 0.15	 & 1.92	 & 	1.19 & 		43.7 &  [3] \\
RXJ1119.7+2126	& 169.899792	& 21.455056	& 2.5   &  /	& -21.47	& 0.06	& 0.21	&	0.57*	&	0.9	&  [4] \\
BLOX J1230.6+1113.3 ID & 187.602827  & 	11.189658 & 2.1	 & 3.5	 & -22.51 & 0.12 & 0.18	  & 0.54 & 3.3 &  [1] \\
XMMXCS J123338.5+374114.9 & 188.410417 & 37.687472  & 2.6	 & 3.2 & -22.44 & 0.10 & 0.22 & 0.58 & 5.6 &   [1] \\
FG12  & 191.713382 & 0.2970410 & 2.0	 & / & -23.62 & 0.09 & 1.73 & 1.15 & 11.0 &  [5] \\
RXJ1331.5+1108	& 202.873551&	11.132486& 2.0  &   /	&	-22.48&	0.08&	0.44	 &	0.73* &		3.3	&  [4] \\
XMMXCS J134825.6+580015.8 & 207.106667 & 58.004389 & 2.0	 & 2.6	 & -23.23 & 0.13 & 0.54	  & 0.78 & 17.8 &   [1] \\
FGS20 & 212.517450 & 41.755800 & 2.17 & 2.46	 & -23.56 & 0.09 & 0.46	  & 0.74	 & 9.5	&  [2] \\
RXJ1416.4+2315	& 214.112083	& 23.258611	& 2.4 &    /	&	-24.05	& 0.14	& 4.41	& 1.57* &		127.8 &  [4] \\
XMMXCS J141657.5+231239.2 & 214.239583 & 23.210889 & 2.8	 & 3.1 & -22.94 & 0.12	 & 0.20	 & 0.56 & 3.2 &   [1] \\
RXJ1552.2+2013	& 238.051250 & 	20.229167 & 2.3 &    /	&	-23.66  &	0.14 &	1.78	&	1.16 &		36.5 &  [4]	 \\
AWM 4 & 241.237500 & 23.920556 & 2.23 & / & -23.07 & 0.03 & 1.33 & 1.05 & 1.5 &  [6] \\
\hline
\end{tabular}
\end{center}   
\tablefoot{Columns represent: (1) Cluster name. (2): Right ascension of the BCG. (3): Declination of the BCG. (4): Magnitude gap between the two brightest member galaxies. (5): Magnitude gap between the first and the fourth brightest member galaxies. (6): Absolute magnitude of the BCG, computed using SDSS DR16. (7): Redshift of the BCG. (8): Virial mass of the object estimated from X-ray data. (9): Virial radius, estimated from X-ray data. (10): X-ray  bolometric luminosity.  (11) Publication from which the original L$_X$ data were taken: [1] \citet{Harrison2012}, [2] \citet{Zarattini2014}, [3] \citet{Proctor2011}, [4] \citet{Jones2003}, [5] \citet{LaBarbera2012}, and [6] \citet{Zibetti2009}.

*R$_{200}$ radius computed in this work from the X-ray luminosity given in the original work. We use equation 2 from \citet{Bohringer2007} to compute R$_{500}$ and then convert it to R$_{200}$ using R$_{200} = 1.516 \times$ R$_{500}$ \citep{Arnaud2005,Girardi2014}}
\end{table*}   

\subsection{FGS28}
\label{sec:FGS28}

FGS28 is found at $z=0.032$. There are four other clusters nearby: NGC 6107 ($z=0.0315$), A2192 ($z=0.0317$), A2197 ($z=0.0301$), and A2199 ($z=0.0299$). The largest velocity difference between all of them is $\sim 600$ km s$^{-1}$. The pair A2197 and A2199 is considered as a supercluster in the literature \citep{Rines2002} and it is known to be connected with a large filament to the Hercules supercluster \citep{Ciardullo1983}. The A2197 mass profile (the closest to the position of FGS28) is better fitted by dividing the cluster in two clumps, that are named East and West in \citet{Rines2002}.

Moreover, in \citet{Zarattini2014} we found that FGS28 was peculiar, since it has only one member within R$_{200}$ and there were other 4 members outside this area. Our interpretation is that this is not a real group of galaxy, whereas it is a giant galaxy that is part of the A2197/A2199 supercluster.

For these reasons, we prefer to remove FGS28 from our sample of FGs.

\section{Fossil systems position in the large-scale structure}
\label{sec:lss}
In this section we discuss how we define the large-scale structure around our FGs. In particular, in Sect. \ref{sec:filaments} we introduce the  catalogue of filaments presented in \citet{Chen2016} and we compute the distance of the FGs in our sample from the centre of filaments and intersections. 

Then, in Sect. \ref{sec:fof} we explain in details how we apply a friends-of-friends algorithm to our FGs and which are the useful output of the algorithm itself.

\subsection{Catalogue of filaments}
\label{sec:filaments}
The first method that we use to study the large-scale structure of our FG sample is to analyse their position with respect to the catalogue of filaments presented in \citet{Chen2016}. In this catalogue, filaments are found using SDSS data by applying the Subspace Constrain Mean Shift (SCMS) algorithm using galaxy density ridges. In particular, SCMS performs three steps to detect filaments \citep[density estimation, thresholding and gradient ascent, see][and references therein]{Chen2015}. To build the catalogue, the authors used spectroscopically-confirmed galaxies in the redshift range $0.05 < z < 0.7$, divided into 130 redshift bins. As a result, the filament catalogue only covers this redshift range and it is limited in RA and Dec (109 $\lesssim$ RA $\lesssim$ 267 and $-4 \lesssim$ Dec $\lesssim 70$).

This approach is similar to that used in \citet{Adami2020}, where they used data from the Canada France Hawaii Telescope Legacy Survey (CFHTLS) to detect FG candidates using photometric redshifts and then used a catalog of filaments and nodes obtained from the same CFHTLS to study the position of their FG candidates with respect to the large-scale structure of the Universe.

\begin{figure*}[ht]
    \centering
    \includegraphics[trim=50 370 50 100,width=\textwidth]{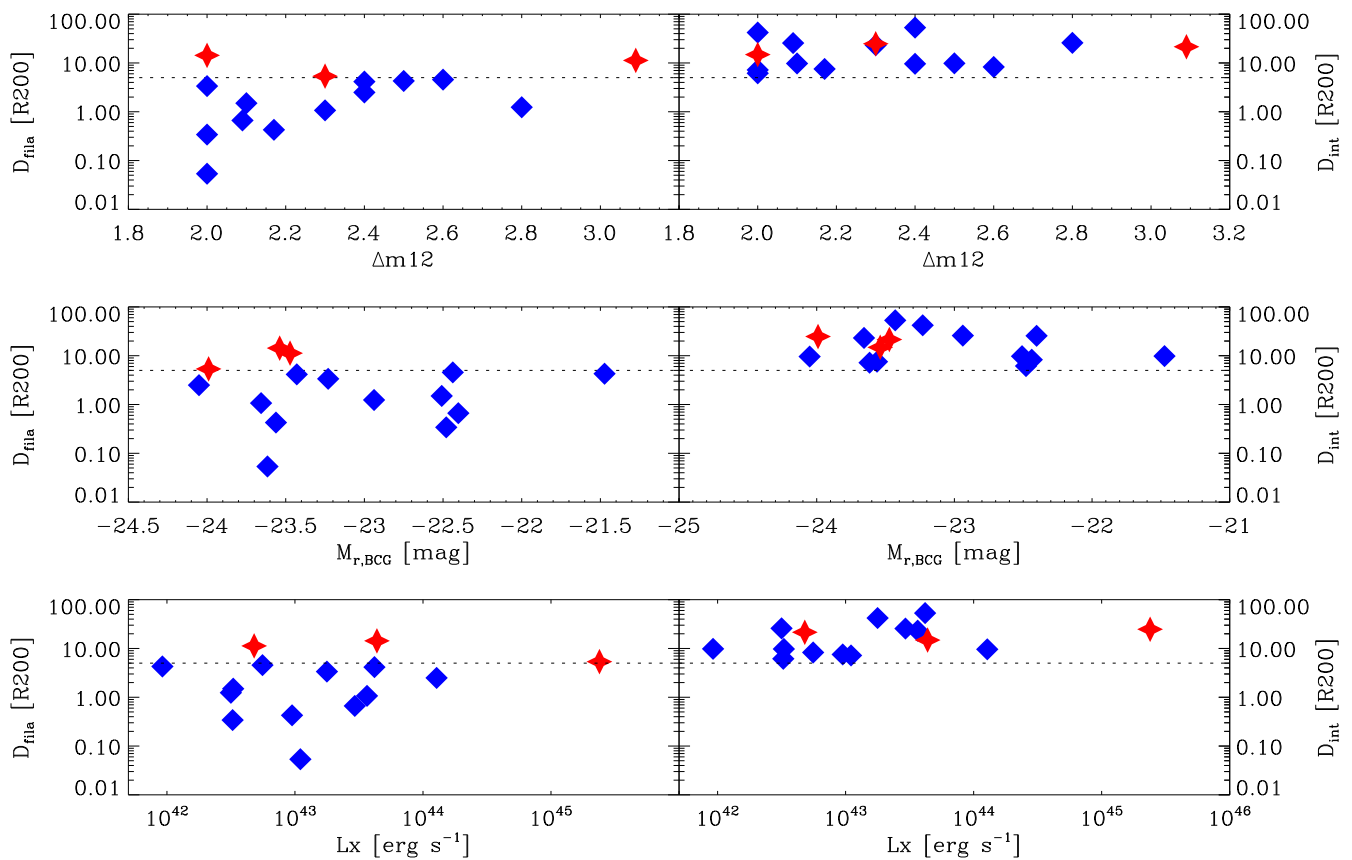}
    \caption{In the left column, correlations between $\Delta m_{12}$ (top panel), M$_{r,BCG}$ (middle panel), and L$_X$ (lower panel) and the distance to the filament in units of R$_{200}$. In the right column, the same quantities are correlated with the distance to the intersections. The dashed horizontal line in the left panel represent 5 R$_{200}$, the distance that we used to separate FGs that are close to filaments ($\rm{D_{fila} < 5}$ R$_{200}$, blue diamonds) from those that are not (red stars).}
    \label{fig:distances}
\end{figure*}

We are now able to measure the distance of our FGs from the centre of the filaments for most of our systems. In particular, we computed the distance in RA, Dec between our FGs and the closest filaments in the redshift space. The filament catalogue gives the minimum redshift of the filament ($z_{low}$) and all the galaxies in the filament satisfy $z_{low} \le z \le z_{low}+0.005$. This means that, in the velocity space, each filament is $\sim 1500$ km s$^{-1}$ wide. 
\begin{table}
\setlength{\tabcolsep}{15pt}
\caption{Distance to filaments and intersections for the FGs in our sample.}
\label{tab:filaments}
\begin{center}
\tiny
\begin{tabular}{lcc}
\hline 
\noalign{\smallskip}   
\multicolumn{1}{l}{Cluster name} & \multicolumn{1}{c}{$D_{fila}$} & \multicolumn{1}{c}{$D_{int}$}\\ 
\multicolumn{1}{l}{} & \multicolumn{1}{c}{[$R_{200}$]} & \multicolumn{1}{c}{[$R_{200}$]} \\
\noalign{\smallskip}  
\hline
DMM2008 IV & 4.1 & 53.1 \\
FGS03 & 0.7 & 25.6 \\
SDSS J0906+0301 & 11.3 & 21.5  \\
A1068 & 5.4 & 24.7	\\
SDSS J1045+0420 & 14.3 & 14.8 \\
RXJ1119.7+2126	& 4.3 & 9.8	\\
BLOX J1230.6+1113.3 ID & 1.5 & 9.8 \\
XMMXCS J123338.5+374114.9 & 4.6 & 8.3  \\
FG12  & 0.05 & 7.2 \\
RXJ1331.5+1108	& 0.3 & 6.1 \\
XMMXCS J134825.6+580015.8 & 3.4 & 42.1  \\
FGS20 & 0.4 & 7.5	\\
RXJ1416.4+2315	& 2.5 & 9.6 \\
XMMXCS J141657.5+231239.2 & 1.2 & 25.8 \\
RXJ1552.2+2013	& 1.1 & 23.1	 \\
AWM 4* & / & / \\
\hline
\end{tabular}
\end{center}   
\tablefoot{Column (1): Cluster name. Column (2): Minimum distance to filament in units of $R_{200}$. Column (3): Minimum distance to intersection in units of $R_{200}$}

* This system has $z < 0.05$ and it is thus outside the redshift interval of the filament catalogue. For this reason, no distance was computed for it.
\end{table}   

We then measured the distance in RA and Dec from the centre of each FG and the centre of the closest filament. We use this definition because in the catalogue of \citet{Chen2016} the position of the centre of the filament (in both Ra and Dec and in the redshift space) is given. We only use filaments that are found within $z \pm 0.005$ from the target, that is again equivalent to $\pm 1500$ km s$^{-1}$, as above. We repeat the same computation for determining the distance of each FG to the closest intersection, that is also provided in the \citet{Chen2015} catalogue. 
In Fig. \ref{fig:distances} we show the relation between the magnitude gap ($\Delta m_{12}$), the X-ray luminosity ($L_X$), and the absolute magnitude of the central galaxy (M$_{r,BCG}$) and the distance from the center of the filaments (left panels) or intersections (right panels), as defined in \citet{Chen2016}. It is interesting to note that the majority of the FGs in our sample are found nearby filaments, with a mean distance of 3.7 $R_{200}$ and a minumim distance of 0.05 $R_{200}$. On the other hands, the distance between our FGs and the intersections is larger, with a mean of 19.3 $R_{200}$ and a minimum distance of 6.1 $R_{200}$, as it shown again in Table \ref{tab:filaments}. In the same table the distance between each FG and the center of the closest filament is reported, but in the Appendix \ref{appendix} we show the entire large-scale structure around each FG for an easier visual inspection.

It is worth noting that for AWM4 we were not able to compute the distance from filaments and intersections, since this FG has $z=0.0317$, a value below the redshift limit of the \citet{Chen2016} catalogue ($z=0.05$).

We are thus able to split our FGs in two categories: systems that are found in a region close to a filament ($D_{fila} \le 5 R_{200}$) and systems that are more isolated. The separation is shown in Fig. \ref{fig:distances} with an horizontal line and in the entire paper FGs close to filaments are shown in blue, whereas red represents FGs that are far from filaments. AWM4, for which the distance to filaments and node was not computed, is eventually shown in black.

However, no correlation is found between $\Delta m_{12}$, $L_X$, and M$_{r,BCG}$ and the distance to filaments or intersections.

\subsection{Friend-of-friends algorithm}
\label{sec:fof}
We also run a friends-of-friends algorithm on our data. The algorithm was presented in \citet{Calvi2011} and was applied to all the galaxies within $\pm 3000$ km s$^{-1}$ from the velocity of the parent cluster. 

\begin{figure*}[ht]
    \centering
    \includegraphics[trim=50 350 50 70,width=\textwidth]{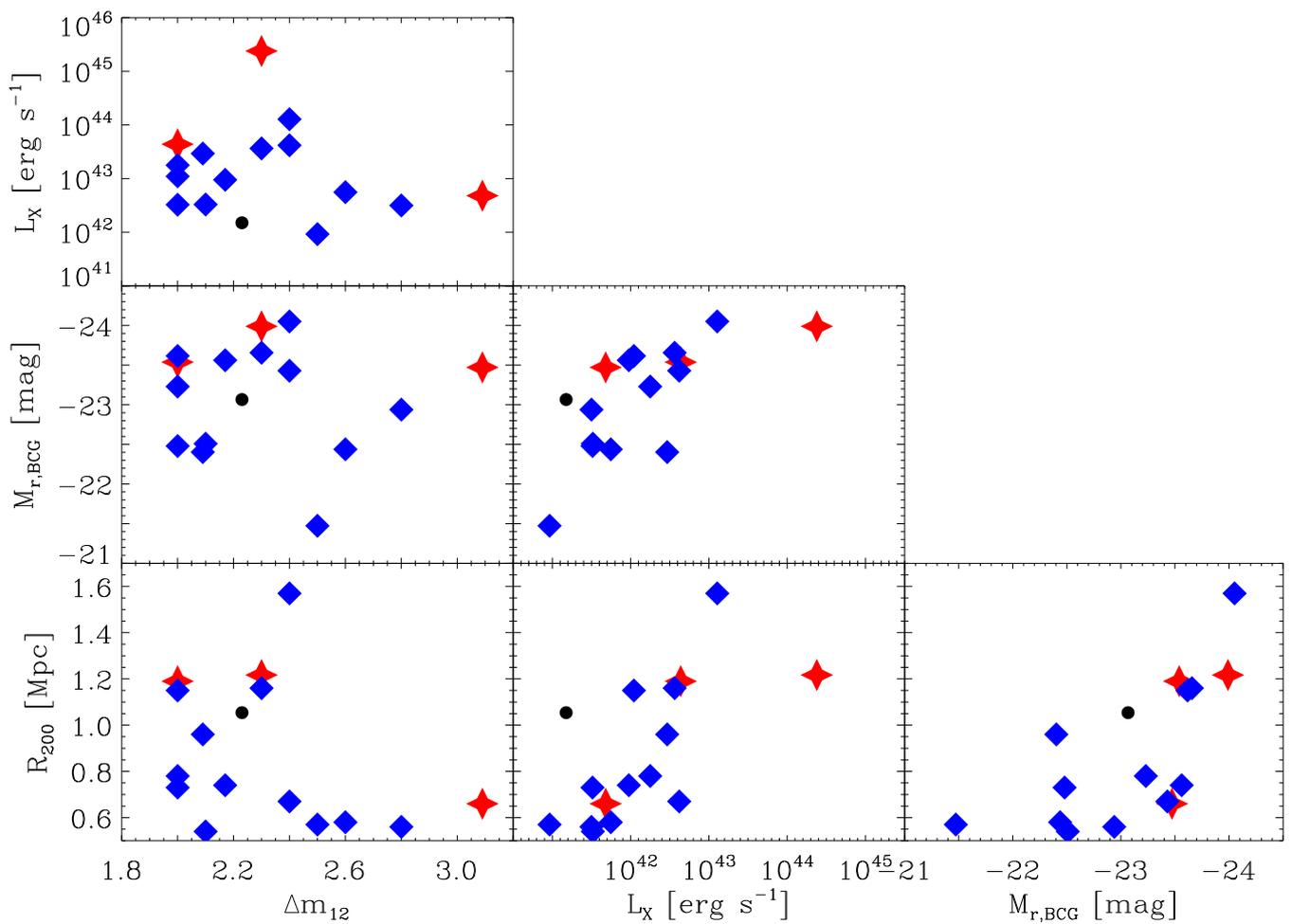}
    \caption{Comparison between the main general properties of the clusters in our sample. The color code is the same as in Fig. \ref{fig:distances}, but here we also show AWM4 as a black dot, since this FG is found at a redshift where the \citet{Chen2016} catalogue is not available and thus it can not be classified according to its distance to a filament.}
    \label{fig:general_properties}
\end{figure*}

Two parameters are used to define friends: a linking length and a linking velocity. For the former, as a first attempt we run the FoF algorithm by using a single linking length of 0.5 Mpc in order to select the core of the clusters and reduce the number of contaminant galaxies. However, the results worsened with $z$: at low redshift ($z < 0.1$) the FoF output follows the visible large-scale structure of each system, but at $z > 0.1$, where data are less sampled, almost nothing was found by the algorithm. Thus, we used a variable linking length depending on redshift: in particular, we used 0.5 Mpc for $z < 0.05$, 1.0 Mpc for $0.05 \le z \ < 0.1$, and 1.5 Mpc for $z \ge 0.1$. This choice was done in order to reflect the smaller number of galaxies found when increasing the redshift. In fact, SDSS spectroscopy is limited in apparent magnitude (down to $r \le 17.77$), which means that increasing redshift will turn in decreasing the number of targets. As a result, the mean distance between galaxies with spectroscopy is expected to grow. On the other hand, the velocity link was chosen to be constant and fixed at $\pm 1500$ km s$^{-1}$. This choice is motivated by the fact that the majority of clusters have velocity dispersions between 300 and 1000 km s$^{-1}$ \citep{Munari2013}. We thus assume that $\pm 1500$ km s$^{-1}$ is enough to include the vast majority of galaxy members, without including too many contaminat galaxies.

In the Appendix \ref{appendix}, the results of the FoF algorithm are presented in red. It can be seen that a good agreement between red points and the filamentary structures presented in \ref{sec:filaments} is found. However, an exact measurement of the precision of the match is beyond the scope of this paper and, in the rest of our work, the filament catalogue will be used as the operational definition of the large-scale structure.

Once the friends-of-friends have been detected, the code looks for virialised structures (e.g. clusters) by computing the $R_{200}$ radius and the velocity dispersion of the group/cluster, removing the outliers and repeating the process until the number of members converges. For our analysis, we limit the detection of a structure to agglomeration of galaxies that have at least three members and a minimum velocity dispersion of 200 km s$^{-1}$ that is the mean velocity dispersion of galaxy groups found using the same FoF algorithm in \citet{Calvi2011}. The goal of this cut is to remove smaller structures like galaxy pairs or smaller groups \citep[][found that only 11\% of their groups have $\sigma_V < 100$ km s$^{-1}$]{Calvi2011}, that we think are not useful for our work. After convergence, we estimate the mass of each structure (details in Sect. \ref{sec:mass}). The main goal of this procedure is to estimate the mass of groups and clusters around our sample of FGs, to check if there is a relation between the available mass and some of the FG properties. However, sometimes also the FG is detected and in this case we are able to estimate the mass in this alternative way. The FoF-computed mass of each FG is presented in the Appendix \ref{appendix}.

\begin{figure*}
    \centering
    \includegraphics[trim=50 360 80 90,width=\textwidth]{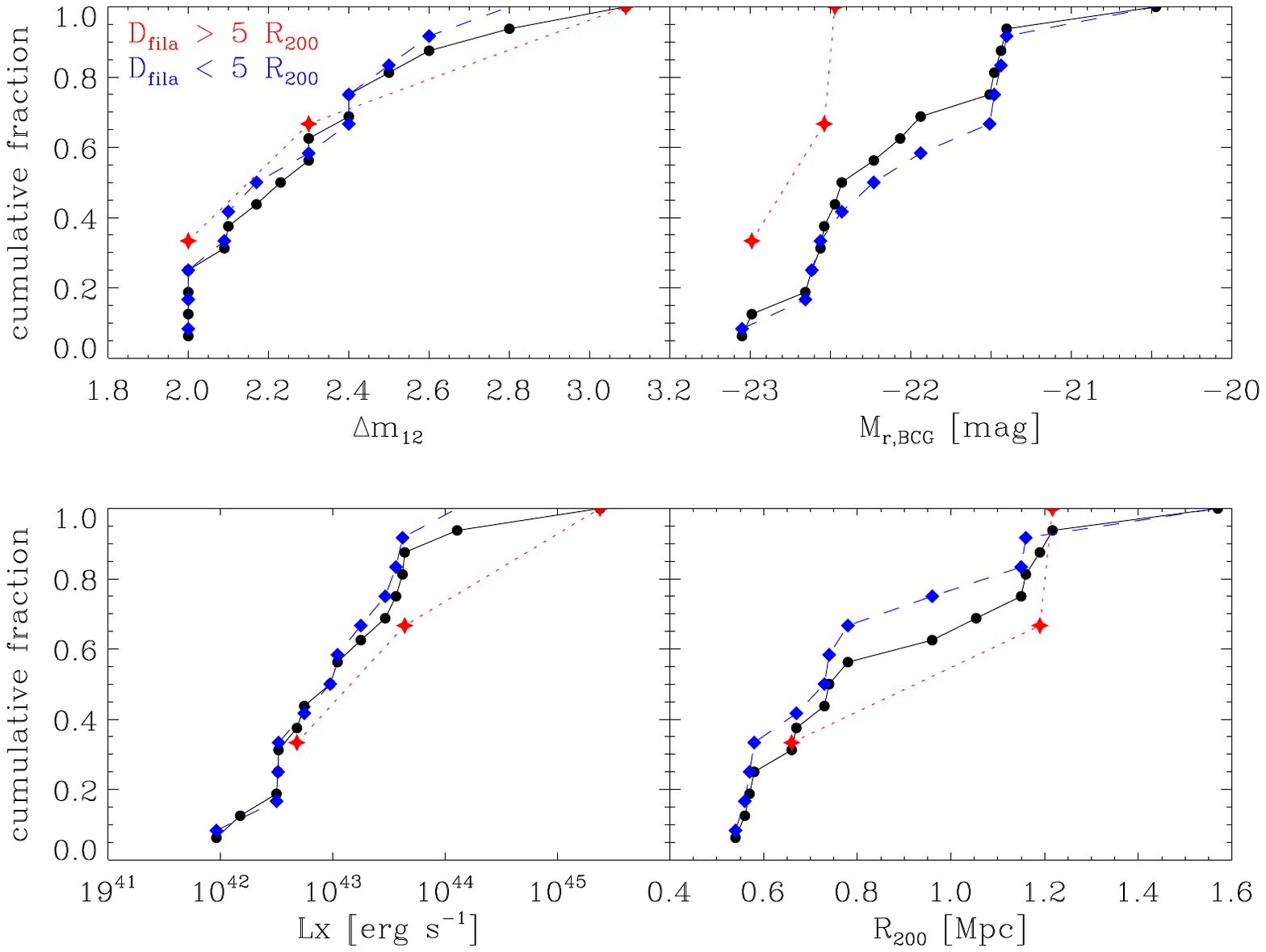}
    \caption{Cumulative distribution of $\Delta m_{12}$ (top-left panel), absolute magnitude of the BCG (top-right panel), X-ray luminosity (bottom-left panel), and $R_{200}$ (bottom-right panel). In each panel, the full FG population is described with the solid black line and black circles, FGs close to filaments are represented with blue dashed line and diamonds, and FGs far from filaments are shown with red dashed-dotted line and stars.}
    \label{fig:cumulative}
\end{figure*}

\section{General properties}
\label{sec:general}

In Fig. \ref{fig:general_properties} we show the correlations between some of the main parameters available for our sample. In particular, we focus our attention on the magnitude gap ($\Delta m_{12}$), the X-ray luminosity (L$_X$), the absolute magnitude of the central galaxy (M$_{r,BCG}$), and the virial radius (R$_{200}$).
Some correlations are visible, such those between $R_{200}$, $L_X$, and the absolute magnitude of the BCG. The $R_{200} - L_X$ correlations is expected, since the former quantity was computed from the latter. Also the $R_{200} - M_{r,BCG}$ and the $L_X - M_{r,BCG}$ correlations are expected for all clusters, since in general massive clusters host massive BCGs \citep{Lin2004,Brough2008}. For our goals it is more interesting to analyse the correlations involving the magnitude gap: it can be seen that there is no clear correlation when using this parameter. The only interesting result is that the systems with the largest magnitude gap ($\Delta m_{12} \ge 2.5$) are small, with $L_X < 10^{43}$ and $R_{200} < 0.7$.

Finally, no specific correlation is found for FGs that are close and far from filaments, although systems with $\Delta m_{12} \ge 2.5$ are mainly close to filaments. However, the one with the largest $\Delta m_{12}$ is far from the closest filament and the statistic is in general very poor for these systems.

We then computed the cumulative distribution of $\Delta m_{12}$, $L_X$, $R_{200}$ and M$_{r,BCG}$ in the two subsamples (e.g. FGs close and far from the centre of filaments). The results, presented in Fig. \ref{fig:cumulative}, show that they follow similar relations, with the exception of the absolute magnitude of the BCG. In fact, the three systems that are far from filaments have all BGCs with $M_r < -23.5$, whereas those close to the centre of filaments are more equally distributed in the range $-24 \le M_r \le -21.5$.

\section{Large-scale mass distribution}
\label{sec:fof_mass}

In this section we want to discuss the large-scale environment of FGs with respect to the quantity of mass available in their surroundings and its distribution. Since SDSS spectroscopic data are not homogeneous, the main issue is how to compute precise areas for all the sample, especially for those FGs which mapping is widely incomplete. For this reason, in \ref{sec:pick} we will discuss the Pick theorem and how we use it for our scopes.
Once the areas are known, we will compute the local projected over density in Sect. \ref{sec:projected}. Finally, in Sect. \ref{sec:mass} we will analyse the quantity of mass available in the FGs' surroundings by using different indicators: the density and cumulative distribution of bright galaxies and the mass found in groups and clusters from the FoF algorithm.

\subsection{The Pick's theorem}
\label{sec:pick}
The statement of Pick's theorem \citep{Pick1899} claims that, if a regular polygon has integer coordinates for the vertices, the area can be computed as

\begin{equation}
    A = i + \frac{b}{2}-1,
\end{equation}
where i the number of integer points inside the polygon and b the number of integer points on its boundaries.

\begin{figure}
    \centering
    \includegraphics[trim=170 350 70 70,width=0.5\textwidth]{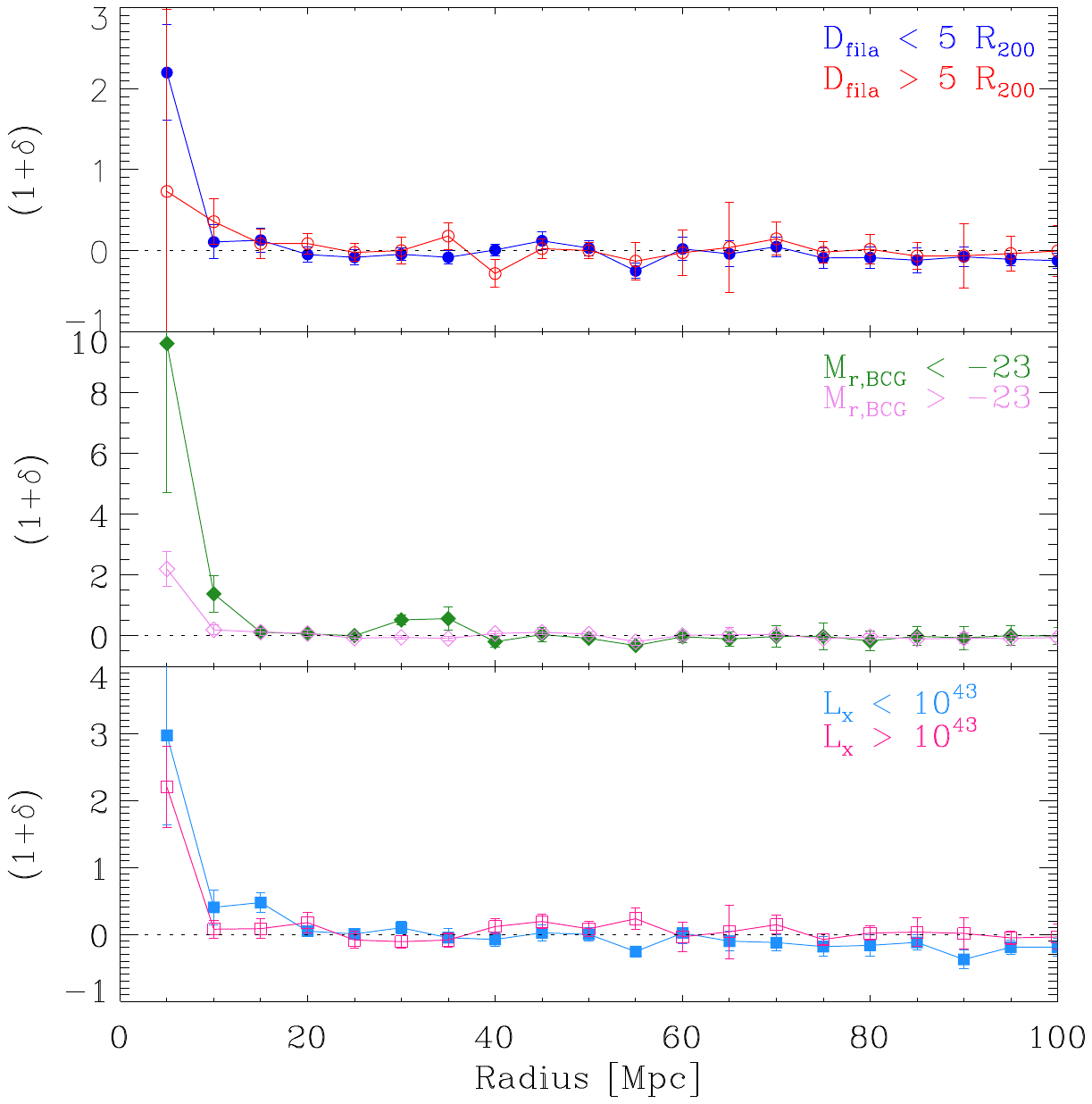}
    \caption{Top panel: galaxy over densities for systems with distance to filament larger and smaller than 5 $R_{200}$ (red open circles and blue filled circles, respectively). Middle panel: same quantities but for systems with bright ($M_r < -23$) and faint ($M_r > -23$) central galaxies (green filled diamonds and pink empty diamonds, respectively). Bottom panel: same quantities but for systems with high ($L_X > 10^{43}$) and faint ($L_X < 10^{43}$) X-ray luminosities (azure filled squares and violet open squares, respectively).}
    \label{fig:overdensities}
\end{figure}

We used this theorem to compute the area available for each FG in the sample. In particular, we divided our areas in different rectangles and we then applied, to these now regular polygons, the Pick's theorem.

To estimate uncertainties, we applied the theorem for those FGs for which the entire 100 Mpc area were available (8 systems). The mean difference between the Pick area and the geometric one is $-2.3$\%, with a maximum of $-3.2$\%. We also compute the difference in the area of 50 Mpc for the 11 systems fully covered out to this radius: in this case, the mean percent error is $-3.2\%$, with a maximum error of $-4.2\%$. The errors are larger in the second case because the number of points used to apply the Pick's theorem is smaller.

We highlight that the differences in measurements are always negative. This means that the Pick's theorem is systematically underestimating the geometric areas. Since the errors are connected to the number of available points, we expect to have larger errors for the farthest FGs, since we are using a sample of galaxies that is limited in apparent magnitude.

\subsection{Local projected over densities}
\label{sec:projected}
We were now able to compute the local over densities within circular coronas around our FGs. In particular, we used the Pick's area for the coronas and only galaxies within $\pm 1500$ km s$^{-1}$ and $m_r\le 17.77$ (the SDSS completeness limit for the main spectroscopic sample). Finally, we corrected for spectroscopic completeness. The over density is computed as 

\begin{equation}
    OD = 1 + \frac{\Sigma (r)-\overline{\Sigma}}{\overline{\Sigma}} = 1 + \delta
\end{equation}
where $\Sigma (r)$ is the density at each specific bin radius and $\overline{\Sigma}$ is the mean density computed by using all galaxies between 20 and 50 Mpc.
In particular, we firstly analyse the over densities for systems that are closer or farther than 5 $R_{200}$ from the centre of the closest filament. No difference is found within the errors in this case, as it can be seen in the top panel of Fig. \ref{fig:overdensities}. Then, we compute the over densities by dividing our sample in systems with bright and faint central galaxies, using $M_r = -23$ as separation. The result is shown in the central panel of Fig. \ref{fig:overdensities} and in this case a difference is found in the very central distance bin, at least at 1-$\sigma$ level. Finally, in the bottom panel of Fig. \ref{fig:overdensities} we plot the over densities of our systems divided using X-ray luminosity, that is a proxy of the mass of our systems. Again, no difference is found for the over densities of small or massive FGs.

\begin{figure*}
    \centering
    \includegraphics[trim=50 390 50 120,width=\textwidth]{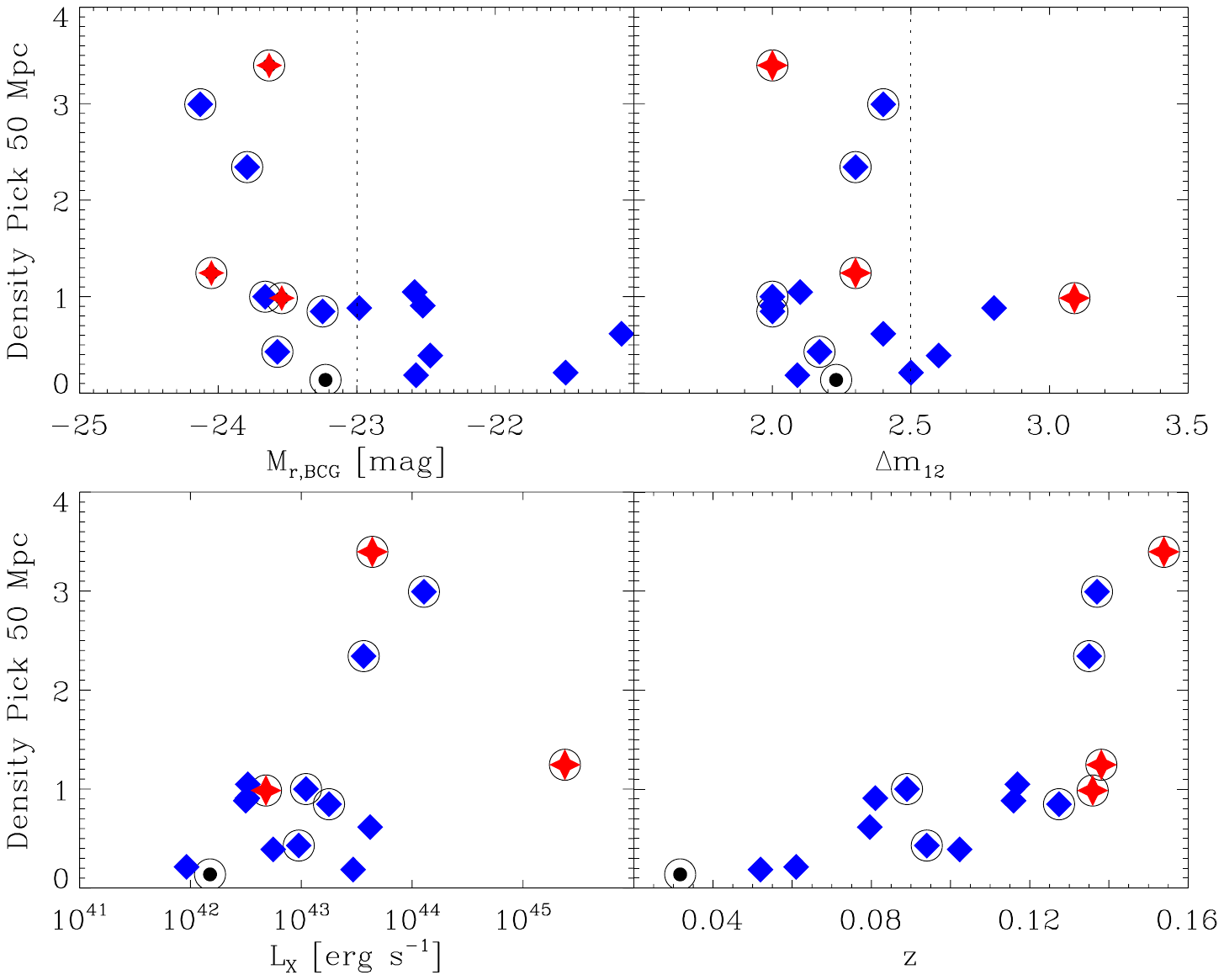}
    \caption{Correlations between the absolute magnitude of the BCG (top left panel), the magnitude gap (top right panel), the X-ray luminosity (bottom left panel), and redshift (bottom right) versus the density of galaxies brighter than -22 within 50 Mpc. The area is computed using the Pick theorem, so footprint incompleteness is properly taken into account. Color code and symbols as in Fig. \ref{fig:distances}. Big open circles represent FGs with central galaxies brighter than -22.} 
    \label{fig:brighter}
\end{figure*}

\subsection{Available mass in FGs' surroundings}
\label{sec:mass}
In order to estimate the mass available in the surrounding of our FGs, we use a set of indicators.

First of all, we compute the density of bright galaxies ($M_r < -22$) per Mpc$^{2}$ that are found within 50 Mpc from the centre of the FG. The area in degrees was obtained using the Pick's theorem, as explained in Sect. \ref{sec:pick}, to have a proper estimate also for systems that are not fully covered in the 50 Mpc radius. In Fig. \ref{fig:brighter} we show the correlations between the absolute magnitude of the BCGs, the magnitude gap, the redshift, and the X-ray luminosities versus the density of bright galaxies. The most relevant result is obtained for the $\Delta m_{12} -$ density correlation: systems with the largest magnitude gap are always found in low-density regions, whereas systems with the smallest gaps are found in all environments. We repeat the computation using galaxies with $M_r \le -23$, finding exactly the same trend, although with less statistics, and for this reason we prefer to show here the results with $M_r \le -22$.

\begin{figure*}
    \centering
    \includegraphics[trim=170 420 200 230,width=0.5\textwidth]{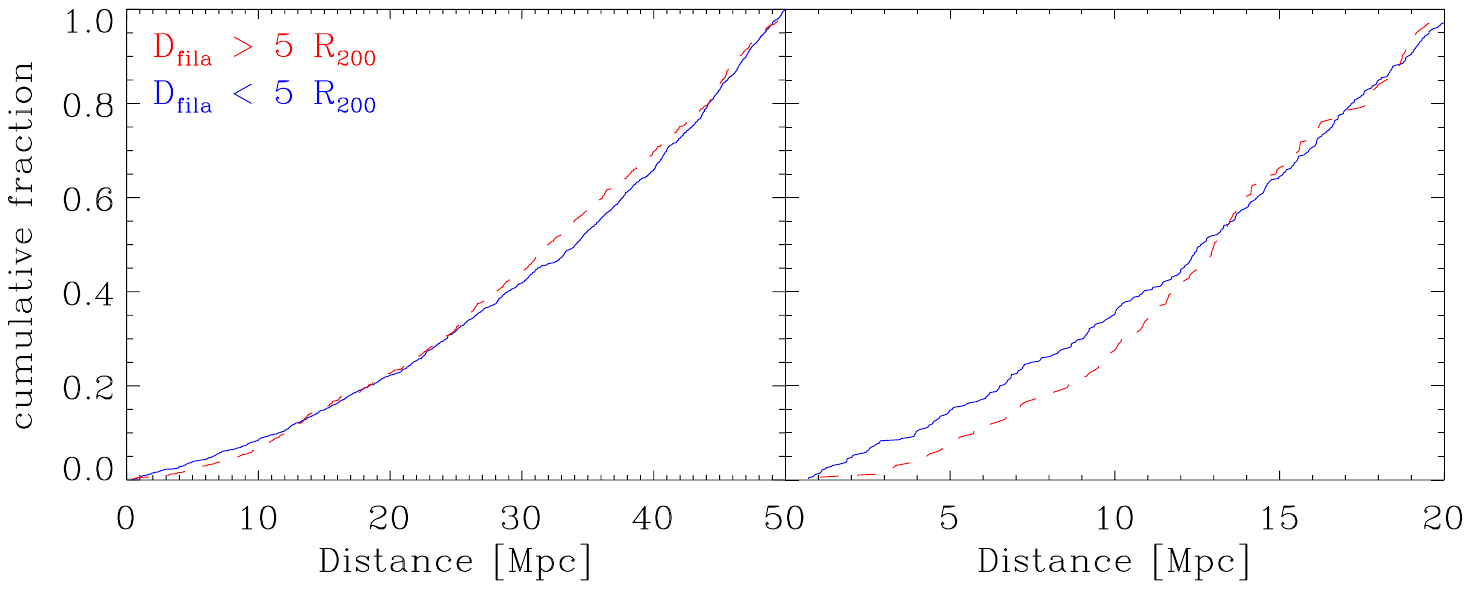}
    \caption{Cumulative distribution of bright galaxies ($M_r < -22$) in 50 Mpc (left panel) and 20 Mpc (right panel). The color code is the same of Fig. \ref{fig:distances}}
    \label{fig:cumulative_bright}
\end{figure*}

Another useful indicator is the cumulative distribution of bright galaxies, that is shown in Fig. \ref{fig:cumulative_bright}. It can be seen that no differences are found within 50 Mpc between FGs close and far from filaments. A Kolmogorov-Smirnov (KS) test confirms this result by giving a KS probability of 0.99, where differences in the two distributions are expected if this value is smaller than 0.05. However, in the first panel of Fig. \ref{fig:cumulative_bright} small differences seem to appear in the central and external regions: in the former, a smaller number of bright galaxies are found in FGs close to filaments, whereas the opposite is found in the latter. We thus compute the cumulative distribution in a smaller area, namely 20 Mpc, aiming at looking in more details to the differences in the neighborhood of our FGs. Although the difference is visually larger, a KS test confirms that both distributions (FGs close and far from filaments) come form the same parent distribution (KS probability of 0.99 also in this case), so we conclude that no difference is found in the distribution of bright galaxies within 50 Mpc from our FGs and, thus, we did not test the external regions.

\begin{figure*}
    \centering
    \includegraphics[trim=70 340 60 80, width=0.95\textwidth]{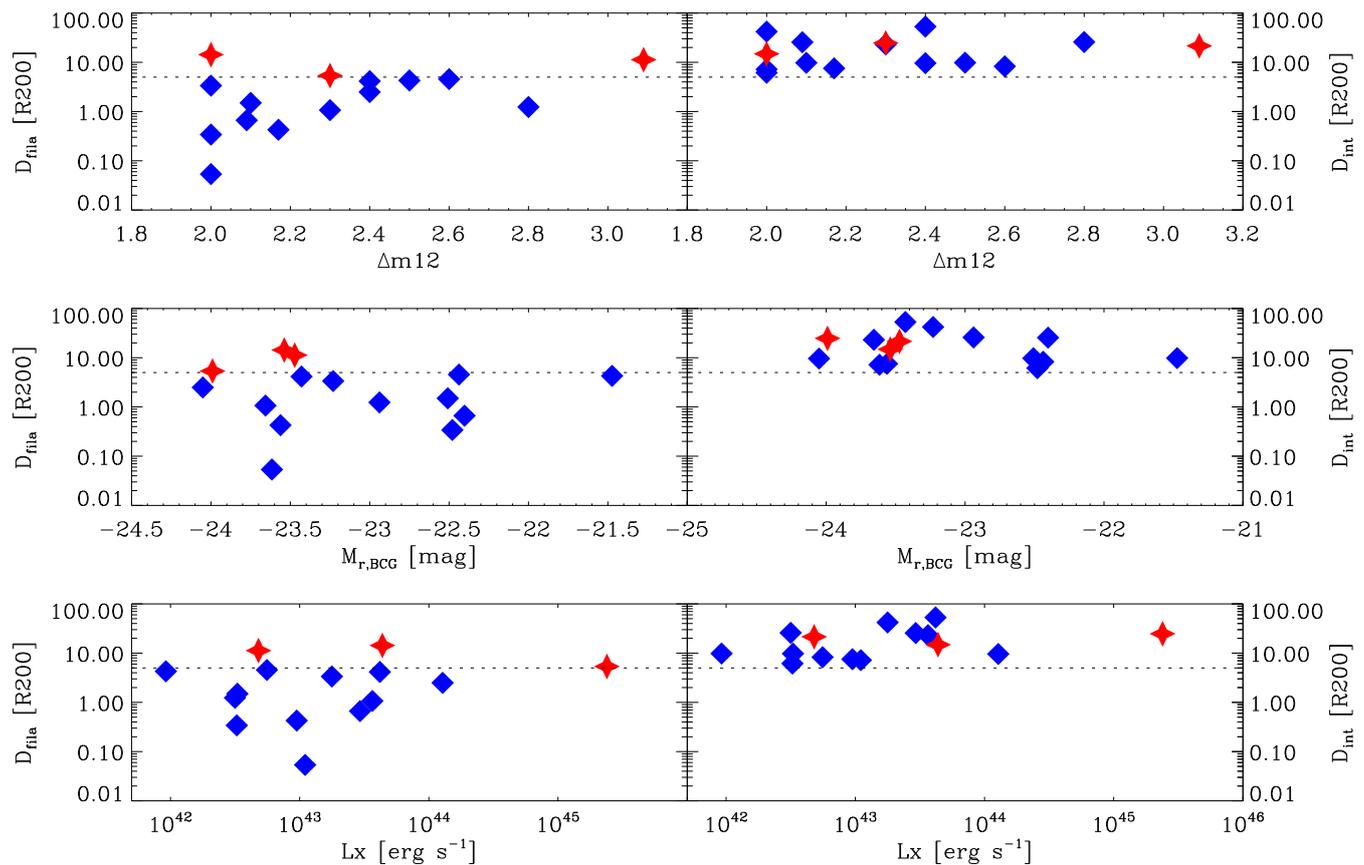}
    \caption{Correlations between the mass found by the FoF algorithm within 50 or 100 Mpc (left and right columns, respectively), and some global quantities of our sample. The color code is the same as in Fig. \ref{fig:general_properties}}
    \label{fig:correlations_mass}
\end{figure*}

The FoF algorithm that we used is able to identify virialised objects and to estimate their velocity dispersion. We thus use it to estimate the mass of these virialised systems according to Eq. 1 of \citet{Munari2013}. In Fig. \ref{fig:correlations_mass} we plot the correlation between total mass available within 50 and 100 Mpc and some global quantities of the FG in our sample, namely the magnitude gap, the absolute magnitude of the BCG, the X-ray luminosity, the virial radius, and the redshift.
There are 11 FGs well mapped out to 50 Mpc and 8 fully mapped out to 100 Mpc. 
No particular trend is found, except the one between the available mass and the X-ray luminosity. In fact, it seems that the most luminous FGs have less mass available, but this result can be driven by a couple of points in the most extreme regions of this relation.  We run a Spearman correlation test \citep{Spearman1904} that found a significant correlation ($\rho \sim 0.02$) between the X-ray luminosity and the available mass within 50 and 100 Mpc. In both cases, there is a negative correlation (the available mass decreases while the X-ray luminosity increases). Other weak correlations are found, one between the virial radius and the mass found in 100 Mpc ($\rho \sim 0.09$) and of between the redshift and the mass found in 50 Mpc ($\rho \sim 0.06$). The latter trend is apparently the opposite of the one found between redshift and the density of bright galaxies, we will discuss in Sect. \ref{sec:discussion} this contradiction in more details.

\section{Discussion}
\label{sec:discussion}

The results presented in this paper can be interpreted in terms of the formation scenario and evolution of FGs. We show that the majority of the FGs in our sample are found close to the centre of filaments, with a mean distance of $3.7 \pm 1.1$ R$_{200}$, whereas none is found close to intersections (mean distance of $19.3 \pm 3.6$ R$_{200}$). This is surprising, since usually galaxy groups/clusters are though to form in the nodes of the cosmic web \citep[e.g.][]{Cautun2014}. These models predicts that the feeding mechanism for galaxy clusters is to receive new objects falling along filaments into the node. The position of FGs, far from the nodes, could thus be the responsible for the formation of their large magnitude gap. In fact, \citet{Ponman1994} firstly suggested that FGs could have been isolated from the cosmic web, thus having enough time to merge all their bright galaxies with the BCG without receiving more bright galaxies from the surroundings. Our results seem to favour this scenario, in which objects that are moving withing the filaments are not able to leave them to reach the nearby FGs. However, it is difficult to estimate the size of filaments, in particular its width or thickness. For this reason we are not able to definitely claim that FGs are found embedded in filaments or just outside them. 

It is interesting to note that our results are in good agreement with what was found in \citet{Adami2020}, where the authors analysed a sample of FG identified using a probabilistic approach in the CFHTLS, finding that FGs seem to reside closely to cosmic filaments and do not survive in nodes. In particular, 87\% of their FGs are within 1 Mpc to a filament, whereas 67\% are farther than 1 Mpc from nodes. We are finding a similar trend, but we only have 33\% of our FGs within 1 Mpc from a filament. This fraction rise to 73\% if we consider systems within 2 Mpc from the centre of the filament. This difference can be due to the different cuts used in \citet{Adami2020} to define filaments and nodes close (in the redshift space) to their target FGs: in fact, since they worked with photometric redshift instead of spectroscopic ones, their filament catalogue uses slices that are thicker than ours, possibly leading to projection effects (e.g. more filaments and nodes in the same projected area with respect to \citet{Chen2016}), as they also noted in their work). A similar conclusion can be reached for the distance from the nodes, with the difference that all our FGs (100\%) are farther than 1 Mpc from nodes, whereas in \citet{Adami2020} they found 67\%. We need to use a distance of 10 Mpc to recover a similar fraction for our sample (60\%). Again, differences in the construction of the catalogues of filaments and nodes or different strategy when analysing the data could lead to these differences that, it is worth noticing, are quantitative but not qualitative.

Within this scenario, it is interesting to note that FGs with the largest magnitude gaps are not massive (all the four systems with $\Delta m_{12} \ge 2.5$ have $L_X < 10^{43}$ erg s$^{-1}$). This can be a boost for the merging timescale of bright galaxies within this systems, since the available mass is smaller (fewer massive galaxies) and the relative velocity between galaxies is smaller too, due to the shallower gravitational potential that groups have with respect to clusters. These two conditions, together with the presence of more radial orbits expected in FGs \citep[as predicted theoretically and, later, observationally confirmed in][]{Sommer-Larsen2005,Zarattini2021} are the main ingredient that reduce the timescale of dynamical friction \citep[see eq. 4.2 of ][]{Lacey1993}. 

\citet{Dariush2010} suggested that only FGs with small central galaxies could be real FGs, intended as systems with older formation than regular groups and clusters that passively evolve without many interactions with the cosmic web. Thus, we can now suggest that these old FGs could be those found aside of cosmic filaments, which peculiar position is able to prevent further accretions. 

Our results support the scenario in which FGs could be a transitional stage in the life of a group/cluster and that the magnitude gap can be reduced when a bright galaxy falls into the FG potential \citep{Aguerri2018,Kim2022}. However, we found that the largest gaps are found in FG with a low density of bright galaxies around. In particular, systems with $\Delta m_{12} \ge 2.5$ has less than one galaxy brighter than $M_r = -22$ within 50 Mpc, whereas five out of ten systems with $\Delta m_{12} < 2.5$ have more than one bright galaxy within their 50 Mpc radius. This could mean that systems with $2.0 \le \Delta m_{12} \le 2.5$ can still be changing their gaps and become non-fossils (because they are closer to $\Delta m_{12} = 2.0$ and they have more bright galaxies nearby), but that FGs with $\Delta m_{12} > 2.5$ are probably less inclined to change the gap to values smaller than $\Delta m_{12} = 2.0$. 

A similar suggestion can be obtained for the most-isolated FGs, since they presents only very bright BCGs and are, on average, quite massive. In this case, the boost in the merging timescale could have been given by the higher available mass for the satellite galaxies, whereas their isolation from both filaments and nodes could have avoided the subsequent arrival of new massive satellites.
However, the paucity of isolated FGs prevents us to reach a robust conclusion on this subsample, since only three FGs are isolated in our sample.

We also analysed the galaxy over density within 100 Mpc. We confirmed that at very large radii ($r > 20$ Mpc) no differences can be found, with the over density that oscillates around zero. 
However, when dividing the sample of FGs into those with bright and faint BCGs, a difference can be found (at $1-\sigma$ level) in the most-central distance bin: systems with brightest galaxies are found in larger over densities. 
This could be connected with a larger number of bright satellites available for merging nearby those massive BCG. On average, FGs with BCGs brighter than -23 have $1.5\pm 1.1$ bright galaxies within 50 Mpc, whereas FGs with BCGs fainter than -23 have $0.6 \pm 0.4$ of these galaxies available. The result is not statistically significant, however we note that the latter systems have all less than 1 bright galaxy within 50 Mpc, whereas the former systems spread the entire range between 0 and 3.4 bright galaxies within 50 Mpc.

However, other correlations appear when comparing the density of bright galaxies with the global properties of our FGs. First of all, a correlation between $L_X$ and the density of bright galaxies is found, where systems with small $L_X$ seem to be more isolated. A Spearman test was run, giving a positive correlation (coefficient of 0.56) and rejecting the null hypothesis (probability of 0.02). 
Moreover, a correlation between redshift and the density of bright galaxies is found, too (Spearman coefficient and probability of 0.85 and $3\times 10^{-5}$, respectively). The first interpretation could be a sort of selection effect due to observations. However, \citet{Verevkin2011} showed that the redshift distribution in the SDSS-DR7 is peaked at $z\sim 0.08$ and then quickly drops at higher redshift. We did not expect this distribution to change in more recent data releases, since the complete main spectroscopic sample (the one that we are using here, limited to $m_r = 17.77$) was released with DR7. Newer SDSS releases have indeed more redshifts, but the target selection is different and focused on high-redshift galaxies \citep[e.g. the BOSS survey,][]{Dawson2013}.

If a selection effect would be present, we would also expect the density to be higher for low-redshift object, that is not the case. In fact, we find the opposite correlation. Our interpretation of this result is that FGs at higher redshift are still found in lively environments, where some major merger can still happen. Indeed, this relation can be seen as an indicator of merger probability, higher for systems at higher redshifts, where more bright galaxies (e.g. more mass) is available. Using numerical simulations, \citet{Kundert2017} showed in the middle-left panel of their fig. 7 that the number of major mergers in FGs is still growing in the redshift range $0.2 \lesssim z \lesssim 0.1$, whereas it stops growing for $z \lesssim 0.1$, a result that can explain what we found in the redshift versus density-of-bright-galaxies correlation.

We also studied the mass available around FGs using the FoF algorithm. In this case, we found an opposite trend with $z$: the mass available (within 50 Mpc) is higher for systems at low redshift. However, we believe that this result can be more affected by biases. In fact, we used different (and arbitrary) linking length for systems with different redshift. This was done to have a good visual agreement between the filaments of \citet{Chen2016} and the FoF galaxies. Since the linking length is smaller for systems at low redshift (0.5 Mpc for $z < 0.05$ versus 1.5 Mpc for $z > 0.1$), we can expect an excess of linked galaxies in the lowest-redshift FGs. Moreover, we are also excluding small groups from the FoF computation, thus again favouring the detection of groups at low redshifts, where the density of data is higher. Finally, the density of bright galaxies is more robust when using a survey limited in apparent magnitude, as the SDSS. In fact, $M_r = -22$ is equivalent, at $z=0.15$ to $m_r = 17.2$, that means that the SDSS spectroscopy (e.g. the galaxies used in this work) is complete also at these redshifts.

Concluding, we found that the very large scale environment (distances larger than 10 Mpc) is not having any role in the evolution of FGs. Hints are found that some difference can be due the the environment at distances smaller than 10 Mpc. This can be due to the presence of filaments, whose mean distance is 3.7 $R_{200}$ (or 3.0 Mpc).


\section{Conclusions}
\label{sec:conclusions}
We analysed a sample of 16 FGs with $z \le 0.15$, for which the magnitude gap was spectroscopically confirmed to be $\Delta m_{12} \ge 2$ and with spectroscopic compeleteness larger than 65\% in SDSS-DR16.

The aim of this work is to test the large-scale environment surrounding FGs and, for this reason, we downloaded all the spectroscopic data available in the SDSS DR16 within a 100 Mpc radius.
Our results can be summarised as follows:
\begin{itemize}
    \item The majority of FGs in our sample is found close to the centre of filaments, with a mean distance of 3.7 $\pm$ 1.1 $R_{200}$ (or 3.0 $\pm$ 0.8 Mpc).
    \item At the same time, all our FGs are found far from nodes, with a mean distance of 19.3 $\pm$ 3.6 $R_{200}$ (or 16.8 $\pm$ 2.6 Mpc).
    \item FGs with the largest magnitude gap ($\Delta m_{12} > 2.5$) are small and not massive ($R_{200} < 0.7$ Mpc and $L_X < 10^{43}$ erg s$^{-1}$).
    \item FGs with the largest magnitude gaps ($\Delta m_{12} > 2.5$) are found in low-density environments.
    \item Only FGs with small magnitude gaps ($\Delta m_{12} < 2.5$) can be candidate to be transitional systems (e.g. systems that can become non-fossil in the near future). In fact, some of them are found in dense environments and new mergers can not be excluded.
    \item The galaxy over density at large scale ($r > 20$ Mpc) varies around the zero value (e.g. there is no over density at such scales).
    \item FGs at higher redshift ($z > 0.1$) have a higher probability of suffering other major mergers, since they are found in denser environment than low-redshift FGs.
\end{itemize}

Our interpretation of these results is that FGs are usually found in a peculiar position with respect to the cosmic web: in fact, they seem to be located close to filaments, whereas galaxy groups and clusters are expected to be found close to nodes. The smaller FGs could be the final end product of group evolution and we do not expect them to evolve anymore. On the other hand, massive FGs at $z>0.1$ could still be evolving, since they are found in denser environment, that we interpreted as having a higher probability to suffer other major mergers, as it is also expected from numerical simulations. Finally, we confirmed that the cosmic web seems to be homogeneous at scales larger than 20 Mpc.

We now plan to apply the same techniques presented in this paper to a larger sample of clusters and groups, spanning the $\Delta m_{12}$ range between $0 \le \Delta m_{12} < 2$, that is complementary to the sample analysed in this publication. In this way, as we already did e.g. in \citet{Zarattini2015}, we will look for dependencies between the position of groups/clusters in the cosmic web and their magnitude gaps.

\section*{Acknowledgements}
We thank the anonymous referee for her/his comments that helps in clarifying the paper and in particular the discussion of the results. SZ is supported by Padova University grant Fondo Dipartimenti di Eccellenza ARPE 1983/2019. JALA was supported by the Spanish Ministerio de Ciencia e Innovaci\'on by the grant PID2020-119342GB-I00. RC acknowledges financial support from the Agencia Estatal de Investigaci\'on del Ministerio de Ciencia e Innovaci\'on (AEI-MCINN) under grant ``La evoluci\'on de los c\'umulos de galaxias desde el amanecer hasta el mediod\'ia cósmico'' with reference PID2019-105776GB-I00/DOI:10.13039/501100011033.

Funding for the Sloan Digital Sky Survey IV has been provided by the Alfred P. Sloan Foundation, the U.S. Department of Energy Office of Science, and the Participating Institutions. SDSS-IV acknowledges support and resources from the Center for High-Performance Computing at the University of Utah. The SDSS web site is www.sdss.org. SDSS-IV is managed by the Astrophysical Research Consortium for the Participating Institutions of the SDSS Collaboration including the Brazilian Participation Group, the Carnegie Institution for Science, Carnegie Mellon University, the Chilean Participation Group, the French Participation Group, Harvard-Smithsonian Center for Astrophysics, Instituto de Astrof\'isica de Canarias, The Johns Hopkins University, Kavli Institute for the Physics and Mathematics of the Universe (IPMU) / University of Tokyo, Lawrence Berkeley National Laboratory, Leibniz Institut f\"ur Astrophysik Potsdam (AIP), Max-Planck-Institut f\"ur Astronomie (MPIA Heidelberg), Max-Planck-Institut f\"ur Astrophysik (MPA Garching), Max-Planck-Institut f\"ur Extraterrestrische Physik (MPE), National Astronomical Observatories of China, New Mexico State University, New York University, University of Notre Dame, Observat\'ario Nacional / MCTI, The Ohio State University, Pennsylvania State University, Shanghai Astronomical Observatory, United Kingdom Participation Group, Universidad Nacional Aut\'onoma de M\'exico, University of Arizona, University of Colorado Boulder, University of Oxford, University of Portsmouth, University of Utah, University of Virginia, University of Washington, University of Wisconsin, Vanderbilt University, and Yale University.  
 
\bibliography{bibliografia}

\appendix
\section{Plots for visual inspection}
\label{appendix}
In this appendix we present plots that are useful for having an at-a-glance view of the large scale environment of each FG in our sample. 

In the following, we will discuss some features of specific FGs that can be deduced from the plots themselves. These are mainly qualitative comments.

\subsection*{DMM 2018 IV}
This FG is found at $z=0.0796$ and SDSS data only map approximately 50\% of the 100 Mpc radius. In particular, the right side (e.g. R.A. $>$ R.A.$_{{\rm FG}}$) is almost complete, whereas very few data are found in the left side. The FG is detected also by the FoF algorithm and we can estimate a mass of $2.7\times 10^{13} M_\odot$ for it. 

Interestingly, various groups are found along the closest filament, plus there is a filament on the right that is not perfectly followed by the FoF groups, although the filament is not completely identified. We suggest that this differences could be due to the incompleteness of the data and that DMM 2018 IV could be close to a node of the cosmic web, although not identified in the \citet{Chen2016} catalogue.

\subsection*{FGS03}
Also for FGS03 the completeness of the 100 Mpc coverage is not 100\%. There is a quadrant, the bottom-right one, that is well mapped and the remaining is less covered in SDSS.

The group is identified in the FoF algorithm and we can estimate the mass of $1.9\times 10^{13} M_\odot$. FGS03 seems to be found within a filament.

\subsection*{SDSSJ0906}
The coverage of SDSSJ0906 is larger than 50\% and it is mainly focused on the upper part (e.g. Dec $>$ Dec$_{{\rm FG}}$). Here the density of galaxies is lower since the redshift is $z=0.1359$, one of the highest of the sample.

However, it is interesting to note that the filaments found on the top-right side of SDSSJ0906 are well followed by the groups and clusters found by the FoF algorithm. This FG is one of the closest to a node.

\subsection*{Abell 1068}
This FG is approximately at the same redshift as the previous one, so similar considerations are valid in terms of number of SDSS galaxies in the region. However, in this case the 100\% Mpc are fully covered by the data.

The system is found to be in a sort of a void, although it is not in the middle but closer to one of the walls.
Many galaxies are found between the FG and this wall, but they are not virialised for the FoF algorithm. This can be interpreted as the presence of another filament in the region, but this is a tentative interpretation and more data are needed to confirm our hypothesis.  

\subsection*{SDSSJ1045}
This FG in the one at the highest redshift in our sample, so again the density of SDSS galaxies is not very high. The coverage of the 100 Mpc is not complete, but it is complete the one of 50 Mpc, that we used for some of our test along the paper.

The FoF algorithm found a small amount of group/clusters, probably due to the low density of point, that reflects in a larger mean distance between point (that is, the linking length of the algorithm). This is why we chose an adapting linking length: in fact, using for example a shorter linking length of 0.5 Mpc nothing was found in this region. Probably, a linking length of 2 Mpc could be better for systems with $0.15 < z < 0.2$, but since this is the only FG in our sample with $z>0.15$ and by a very small amount (in fact, it is at $z=0.154$) we prefer to maintain only three different values for the linking length.

\subsection*{RXJ1119}
The 100 Mpc radius of this FG is fully covered and there are a lot of galaxies in the field, due to the low redshift of the FG ($z=0.06$).

In this case, the effectiveness of the FoF algorithm is very high and a lot of group/systems are found in comparison with other FGs. The identified systems are in good agreement with the position of nodes and filaments from \citet{Chen2016}.

\subsection*{BLOXJ1230}
For this FG ($z=0.12$), the FoF algorithm is finding few groups/clusters. Indeed, the density of galaxies is quite low, despite a full coverage of the 100 Mpc radius.

Most of the FoF detections are on the left side, in a position where a node and 3 or 4 filaments are present. There is still a rather good agreemen between the FoF results and the \citet{Chen2016} catalogue.

\subsection*{XMMXCSJ123338}
For this FG the coverage of the 100 Mpc radius is complete. There are many visible filamnets, expecially in the bottom-left part of the plot, decently mapped also with the FoF algorithm.

However, it is interesting to note that few groups/clusters are found within 50 Mpc from the FG, despite the presence of 6 nodes in the same region, according to \citet{Chen2016}.

\subsection*{FG12}
The sky coverage for FG12 is approximately 50\%: the upper part is well mapped, whereas poor coverage is found in the lower part of the plot.

In this case, the system is also identified from the FoF algorithm, with an estimated mass of $1.1\times 10^{15} M_\odot$.

FG12 is massive and it is found along a filament and very close to a double node. The filamentary stucture is well reproduced also by the FoF algorithm. However, only not-virialised galaxies are found in the closest node.  On the other hand, other quite massive systems are visible just below our FG, so the area seem to be crowded.

\subsection*{RXJ1331}
The sky coverage of RXJ1331 is almost complete on the entire 100 Mpc radius, only a small are in the bottom of the plot is missing.

Galaxies in the central region are recognised as in a structure, but they are not virialised and so we can not say that RXJ1331 is found by the FoF algorithm. For this reason, we are not able to give an estimation of its mass.

This is again a region characterised by a large amount of groups/clusters, especially in the bottom-left part, where a large number of systems are found in a region connecting two nodes. By looking at the FoF results, we also expect a node to be found at approximately R.A. = 188 and Dec = 6, although it is not present in the \citet{Chen2016} catalogue.

\subsection*{XMMXCSJ134825}
Another relatively-high redshift system, with the 100 Mpc radius fully mapped but with low galaxy density.

Something is detected in the centre, but all those galaxies are found to be not virialised, thus we are not able to measure the mass of this FG using the FoF algorithm.

Few virialised objects are found, as for the other high-redshift (e.g. $z > 0.1$) systems of our sample.

\subsection*{FGS20}
In the case of FGS20, the full 100 Mpc area is covered in SDSS and our target FG is also found by the FoF algorithm. We thus estimate for it a mass of $2.3\times 10^{14} M_\odot$ .

There are 6 nodes of the cosmic web within 50 Mpc radius, according to \citet{Chen2016}, but in this case the agreement between this catalogue and our FoF algorithm is not very good.

However, FGS20 is found embedded in a void, closed by filament at all sides, but very close to one of these walls.

\subsection*{RXJ1416}
This is another high-redshift FG for our sample ($z=0.137)$, for which few galaxies are available and the FoF algorithm struggles to find objects, as we already explained for SDSSJ0906.

However, RXJ1416 is found by the algorithm and we can thus estimate a mass of $2.7\times 10^{14} M_\odot$ for it.

This FG is found very close to a filament, but the closest node is at more than 80 Mpc. However, we can not exclude that a node could be found at the position of the FG, since at least four filaments, some of the incomplete, seem to converge to its position. Moreover, we expect a node to be found at 50 Mpc on the left, where a concentration of galaxies are detected as four different systems by the FoF. Also, the absence of nodes on the right and top parts of the plot could be a hint of an incomplete detection in \citet{Chen2016} in this region, as well as of a not-sufficient coverage by SDSS data.

\subsection*{XMMXCSJ141657}
The sky coverage for this FG is complete out to 100 Mpc. The target FG is found by the FoF algorithm, but as a not-virialised structure. We are thus unable to estimate its mass with this method.

A filament is close to our FG, but it seems to be truncated and nothing (filaments or nodes) are found in the central bottom part of the plot. We thus speculate that there could be another undetected filament/node in the region connecting our FGs with the systems found at approximately R.A. = 213 and Dec = 12.
We also speculate that a node could be found close to this FG, due to the apparent convergence of various filaments, some of which not detected in \citet{Chen2016}, like the one that the FoF algorithm seems to find on the left side of the target, almost horizontal.

\subsection*{RXJ1552}
This is another system almost at our redshift limit, with full coverage of the 100 Mpc radius.

Some galaxies are found to be part of a structure at the FG position, but they seem to be not virialised, so we are not able to estimate a mass for our FG in this way.

Filaments and nodes are mostly found in the lower part of the plot, in good agreement with the FoF detections. The majority of the mass in the central 50 Mpc seems to be destined to the node that is found at the bottom-right of the FG.

\subsection*{AWM4}
This is the lowest redshift FG in our sample and its redshift is not compatible with the \citet{Chen2016} catalogue. For this reason, we did not highlight filaments and nodes in this plot.

However, the FoF algorithm seems to find a series of filaments converging to a point that is at the bottom of our FG. Moreover, there are other two possible nodes in the top-right and bottom-left parts of the plot, giving a sort of s-shape to the galaxies detected with the FoF algorithm.

The same algorithm also found a group at the position of our FG, so we can estimate for it a mass of $1.5\times 10^{14} M_\odot$.

It is worth noting that the data for this systems only cover $\sim50$ Mpc radius, nothing is found in the SDSS beyond this limit.

\begin{figure*}
    \centering
\caption{Plots for visual inspections. Black open circles are galaxies with known velocities within $\pm 1500$ km s$^{-1}$. Large open violet circles represent the position of each group/cluster detected by the FoF algorithm (arbitrary radius). Red filled circles are galaxies in these groups/clusters, but not virialised. On the other hand, blue filled circles are galaxies in groups that are also virialised and are used to estimate the velocity dispersion and mass. Coral bands are filaments, whereas light blue big filled circles are nodes of the cosmic web \citep[both from][]{Chen2016}. The red ellipse is centred at the position of the corresponding FG and has a radius of 5 Mpc at the redshift of the target. Green and blue ellipses are centred in the same position, but their radius are 50 Mpc and 100 Mpc, respectively.}
\label{fig:lss-appendix}
\begin{tabular}{@{}c@{}}
    \includegraphics[trim=0 450 0 50, width=0.99\textwidth]{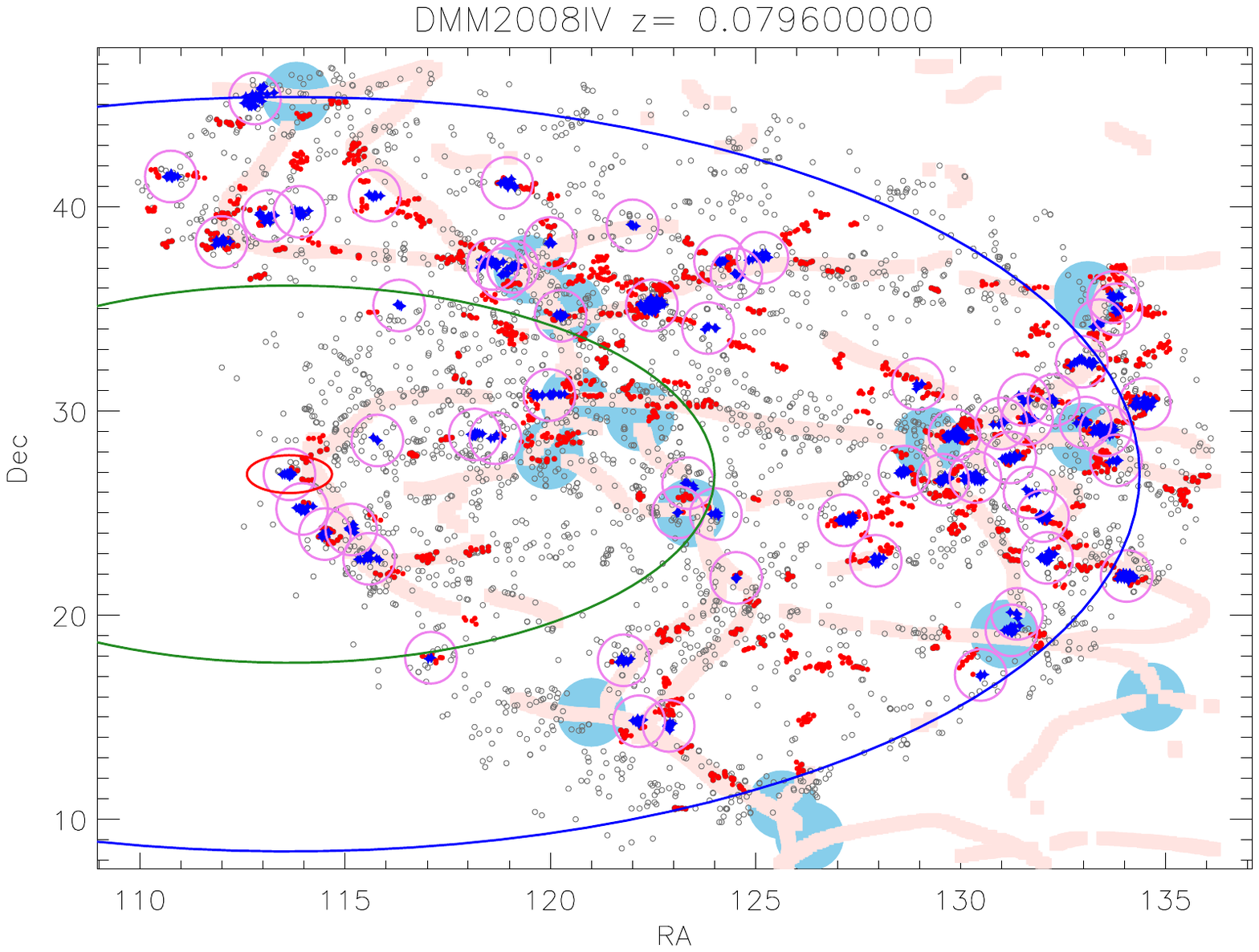}
  \end{tabular}

  \vspace{\floatsep}

  \begin{tabular}{@{}c@{}}
    \includegraphics[width=0.99\textwidth,trim=0 0 0 0]{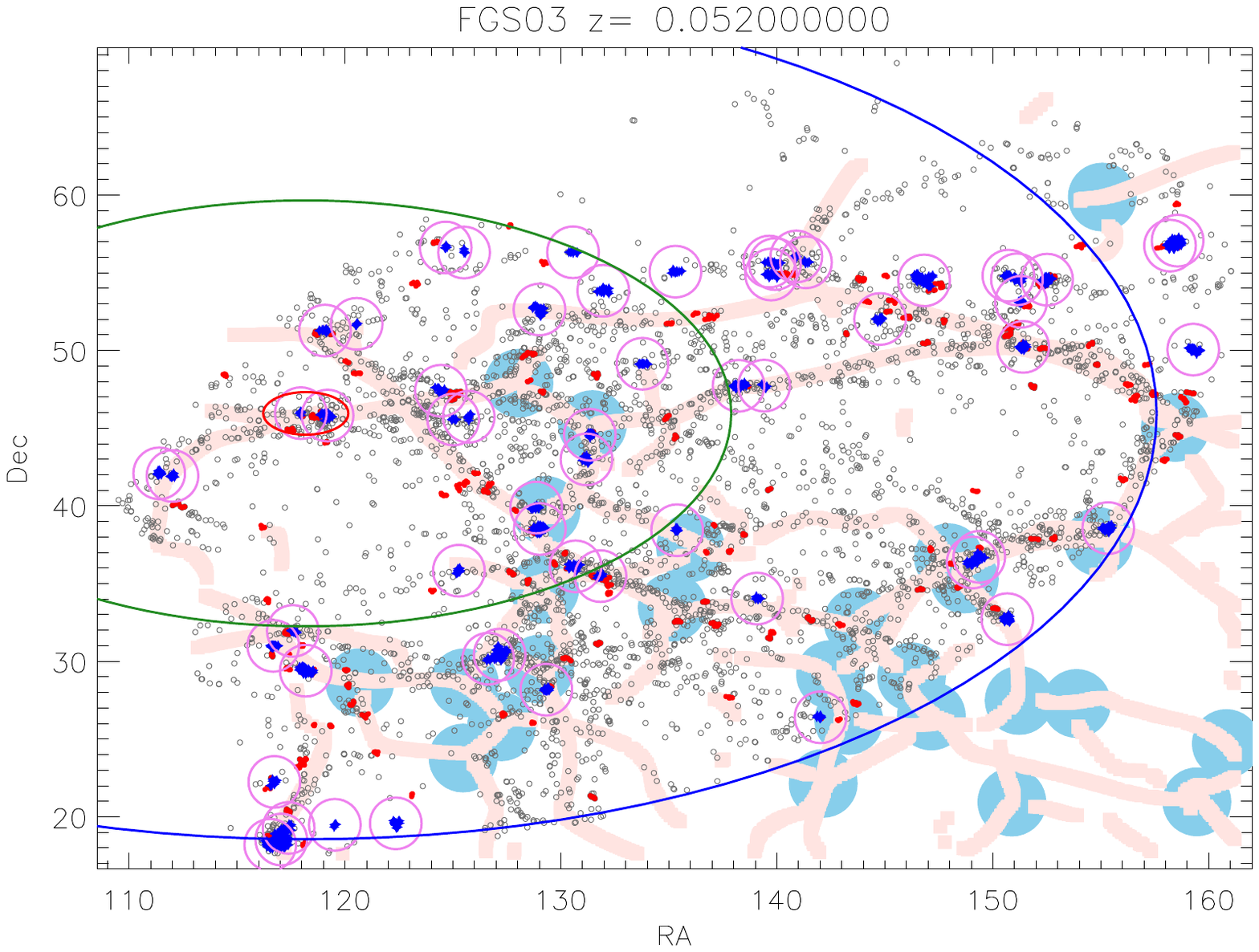}
  \end{tabular}
  
\end{figure*}

\begin{figure*}
\continuedfloat
\caption{Continued}
\centering
\begin{tabular}{@{}c@{}}
    \includegraphics[trim=0 450 0 80, width=0.99\textwidth]{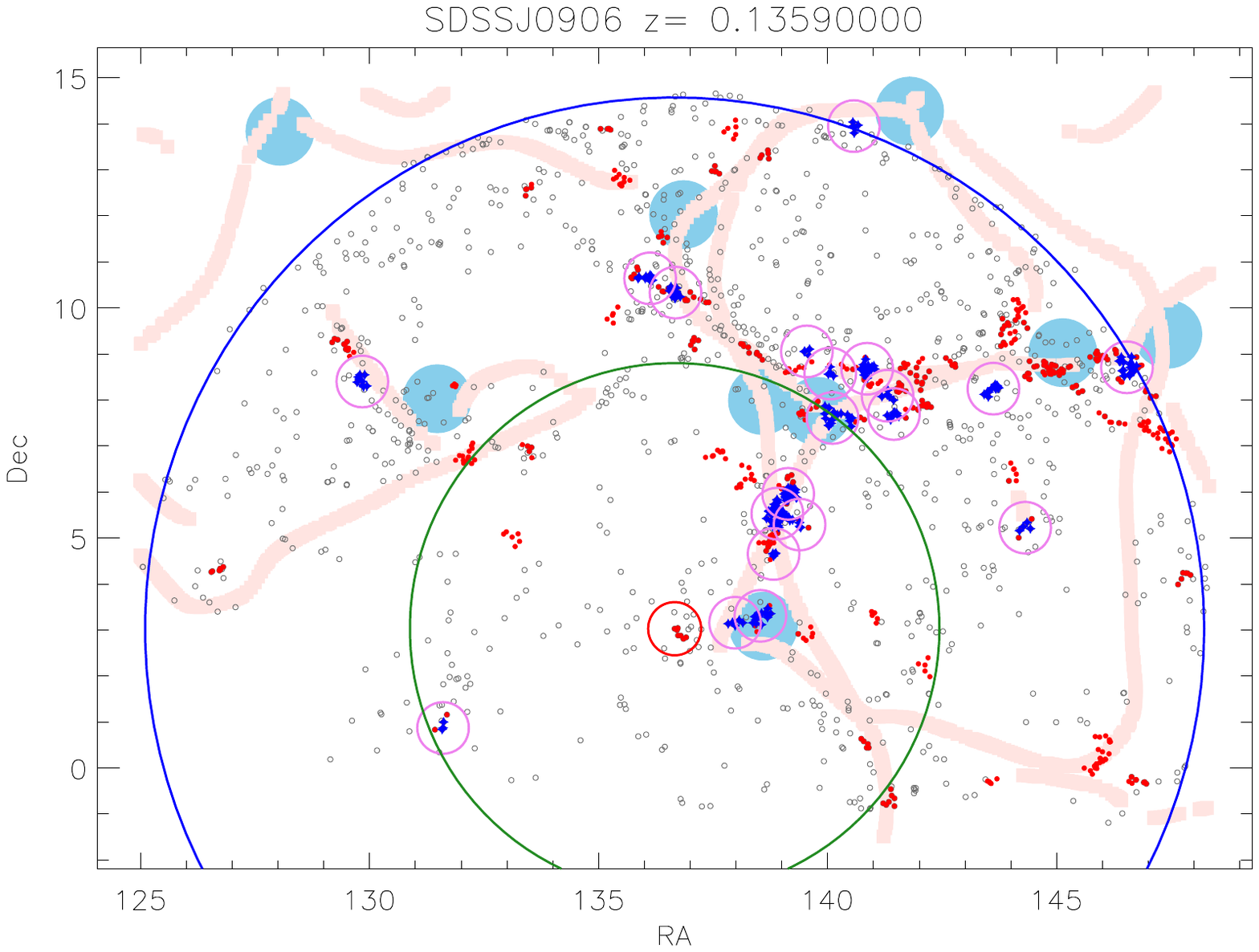}
  \end{tabular}

  \vspace{\floatsep}

  \begin{tabular}{@{}c@{}}
    \includegraphics[width=0.99\textwidth,trim=0 0 0 0]{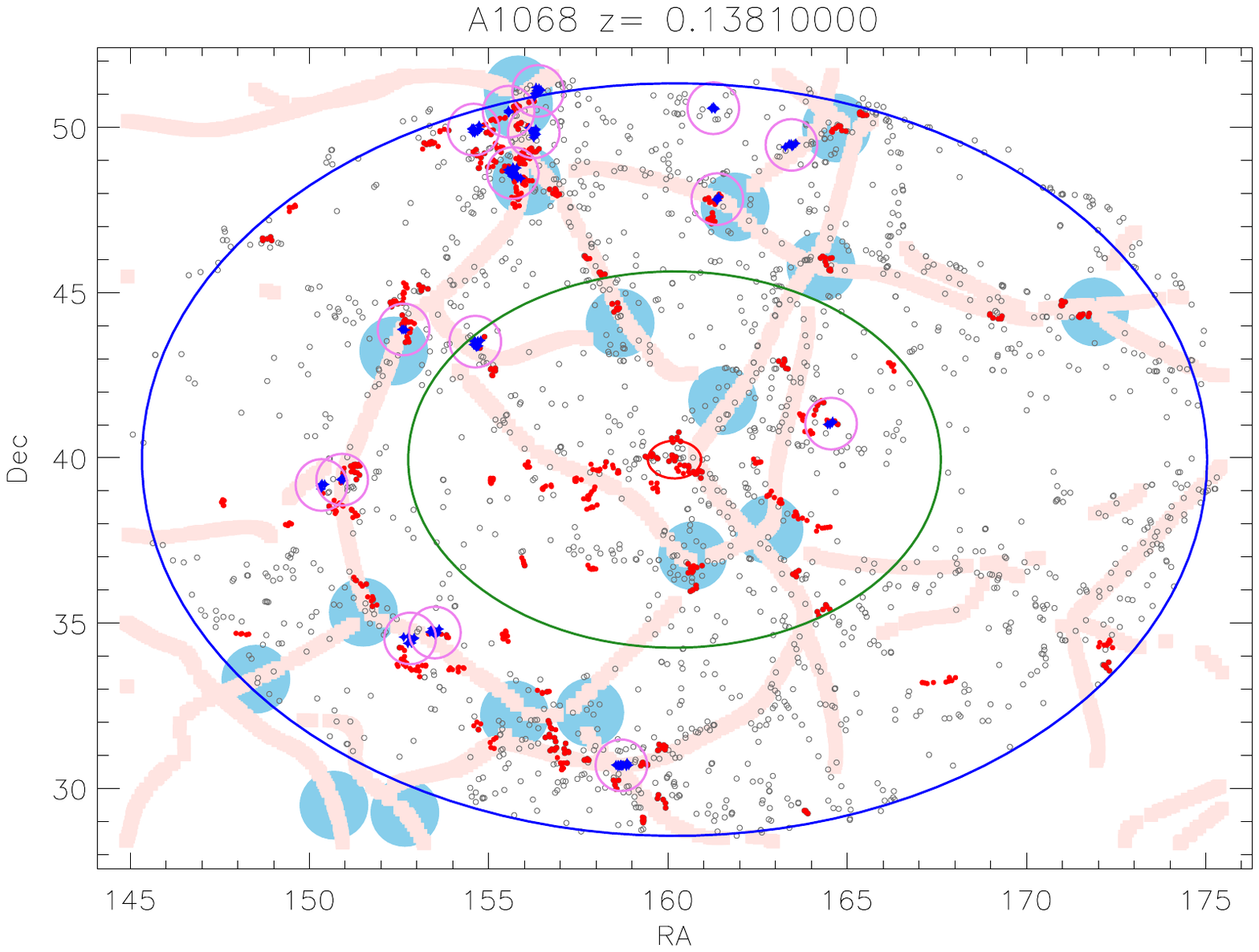}
  \end{tabular}
  
\end{figure*}

\begin{figure*}
    \centering
\continuedfloat
\caption{Continued}
\begin{tabular}{@{}c@{}}
    \includegraphics[trim=0 450 0 80, width=0.99\textwidth]{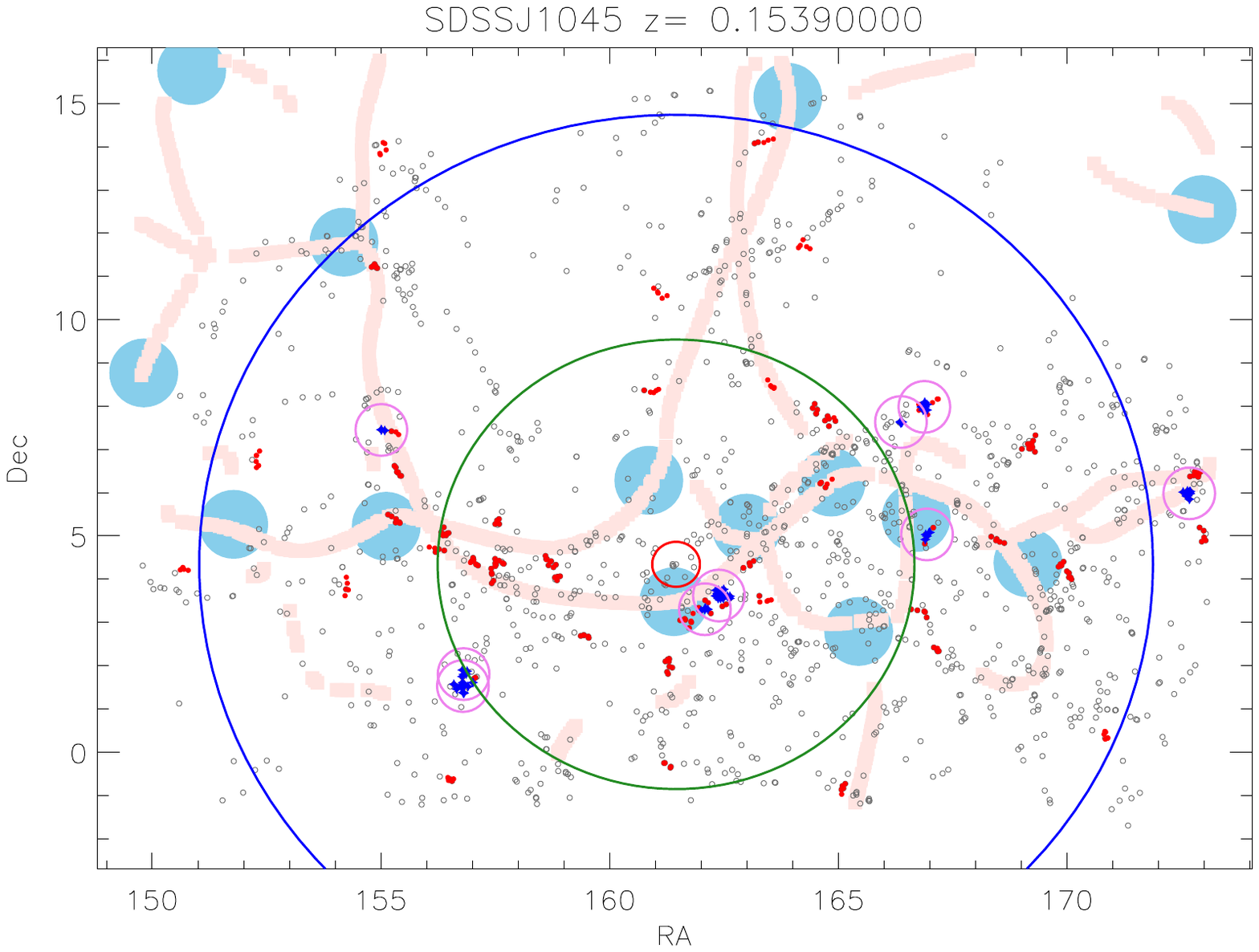}
  \end{tabular}

  \vspace{\floatsep}

  \begin{tabular}{@{}c@{}}
    \includegraphics[width=0.99\textwidth,trim=0 0 0 0]{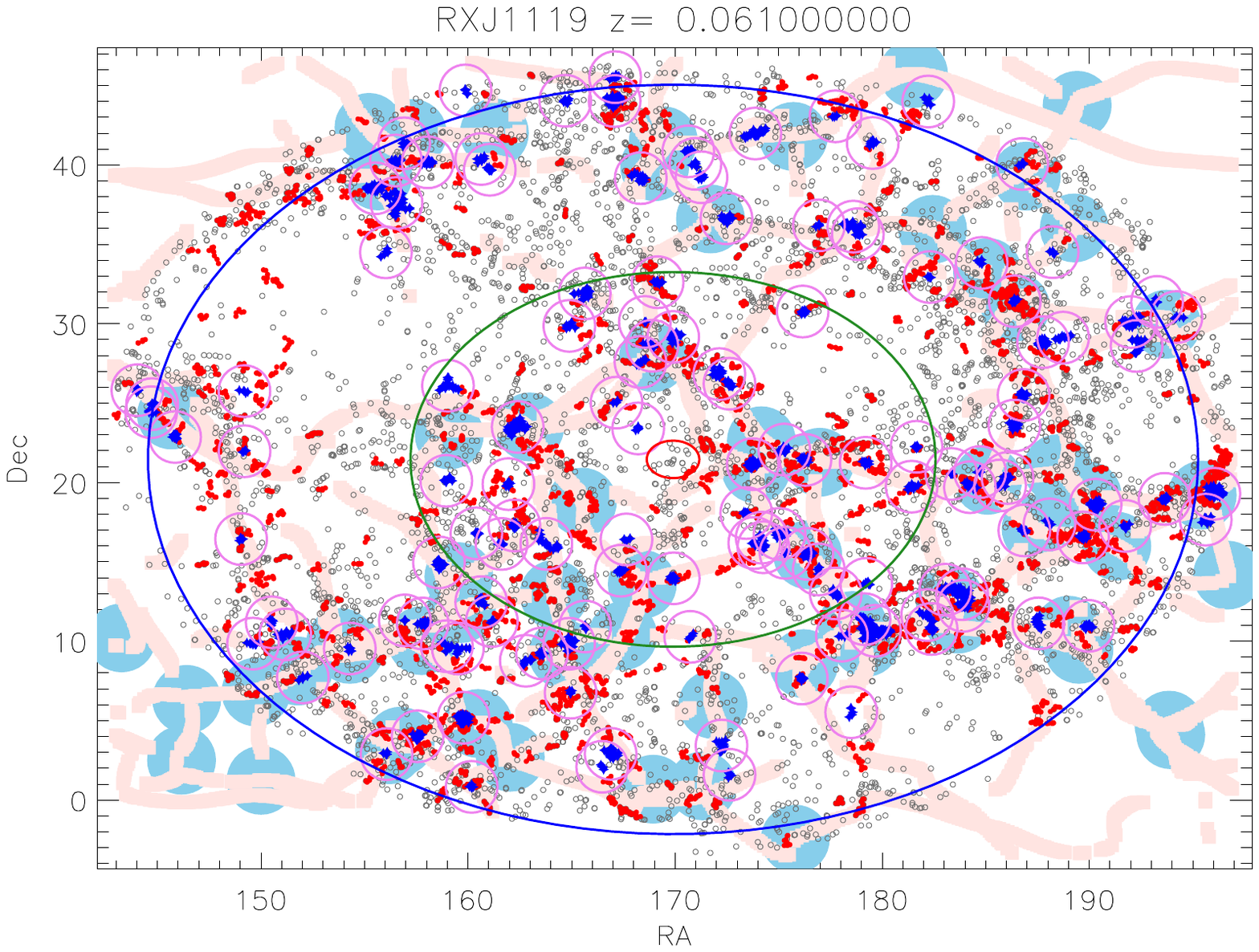}
  \end{tabular}
  \end{figure*}
  
  \begin{figure*}
    \centering
\continuedfloat
\caption{Continued}
\begin{tabular}{@{}c@{}}
    \includegraphics[trim=0 450 0 80, width=0.99\textwidth]{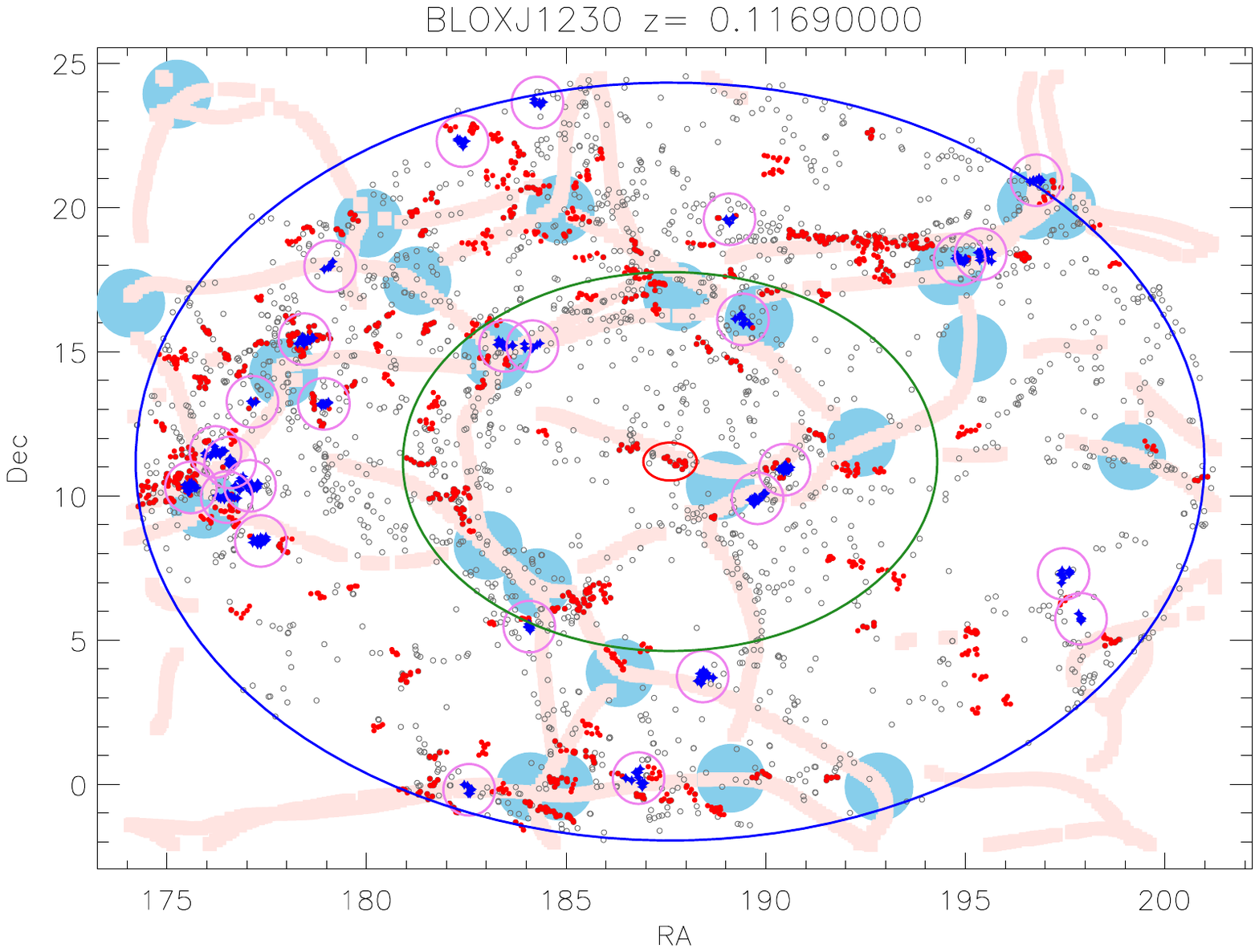}
  \end{tabular}

  \vspace{\floatsep}

  \begin{tabular}{@{}c@{}}
    \includegraphics[width=0.99\textwidth,trim=0 0 0 0]{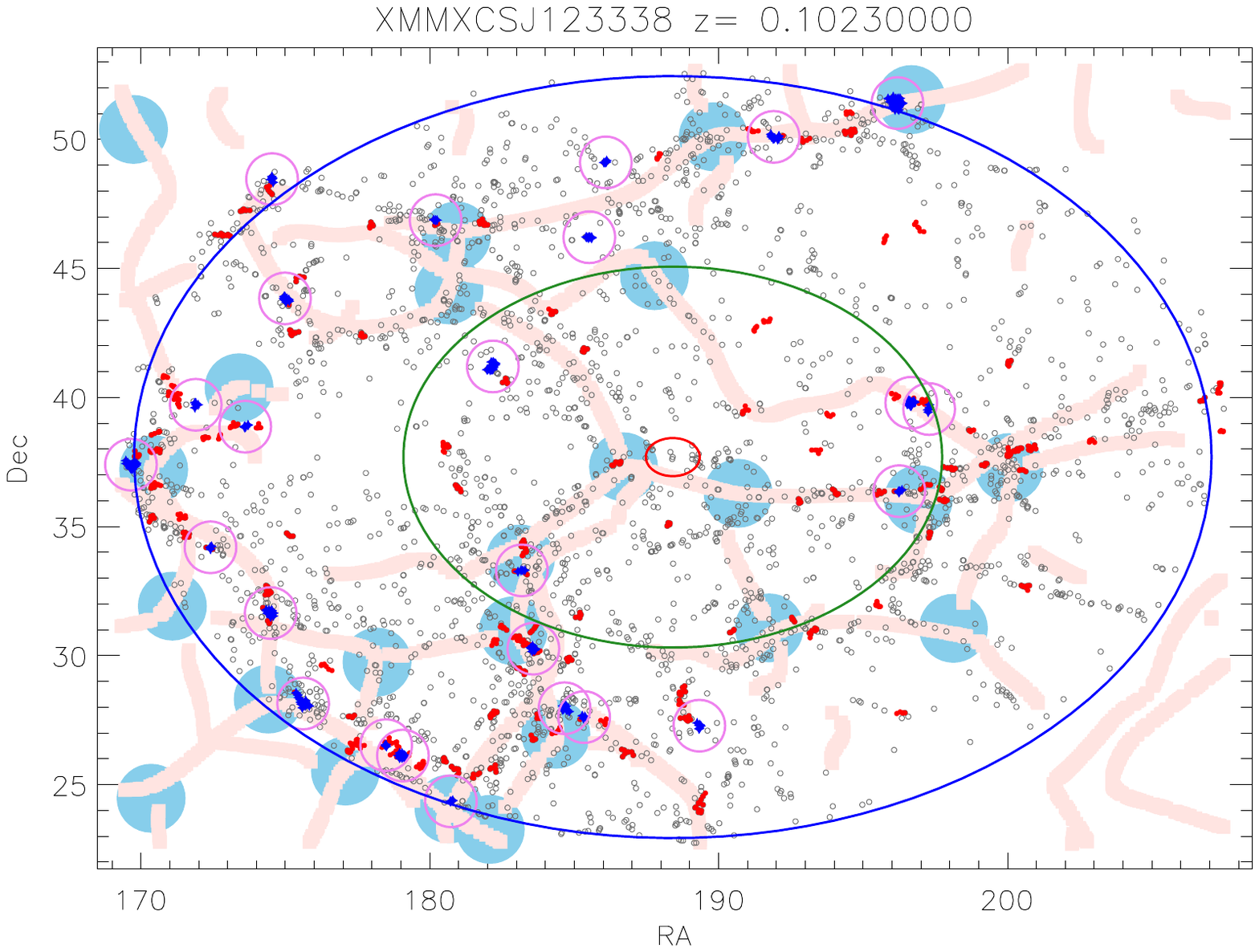}
  \end{tabular}
\end{figure*}

  \begin{figure*}
    \centering
\continuedfloat
\caption{Continued}
\begin{tabular}{@{}c@{}}
    \includegraphics[trim=0 450 0 80, width=0.99\textwidth]{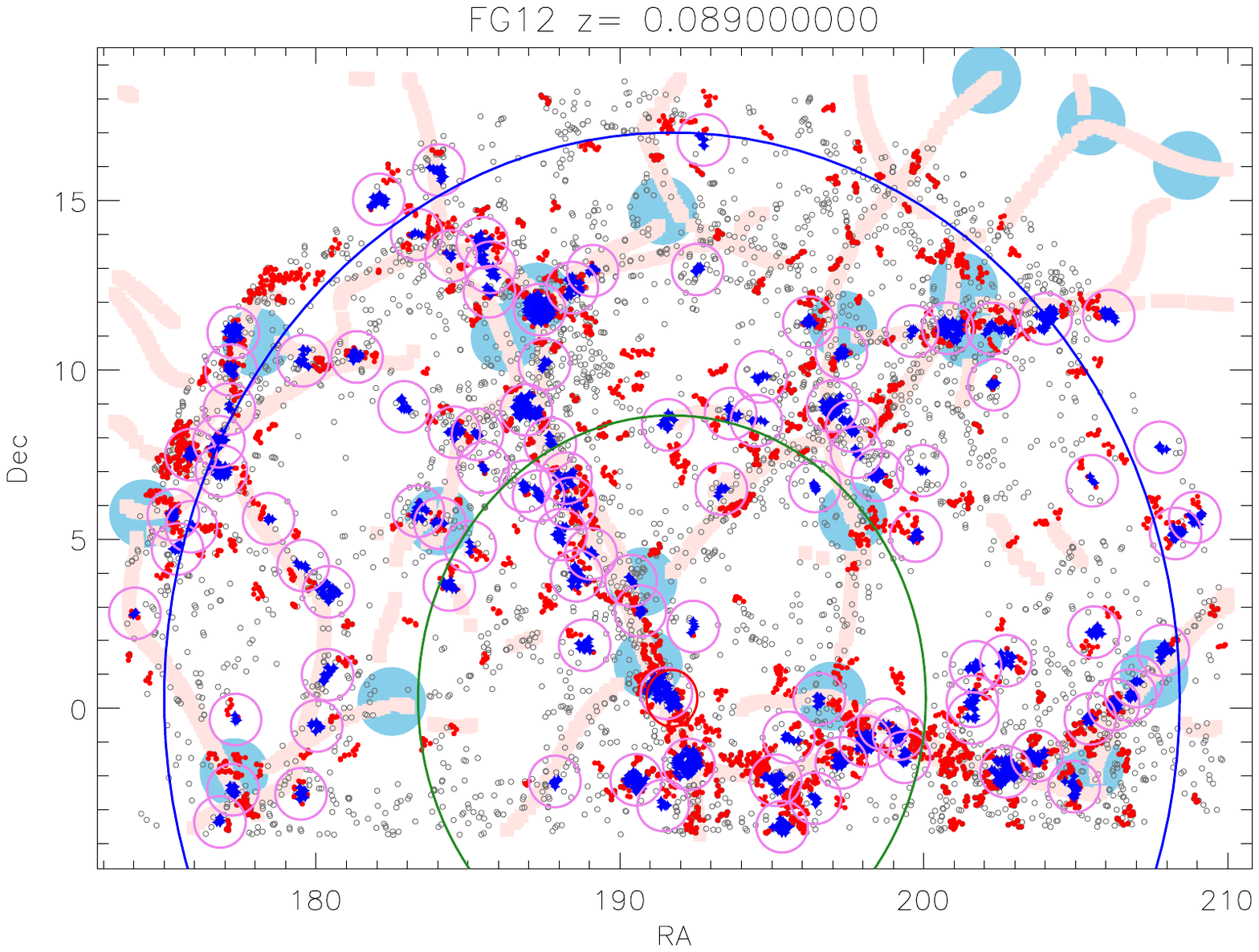}
  \end{tabular}

  \vspace{\floatsep}

  \begin{tabular}{@{}c@{}}
    \includegraphics[width=0.99\textwidth,trim=0 0 0 0]{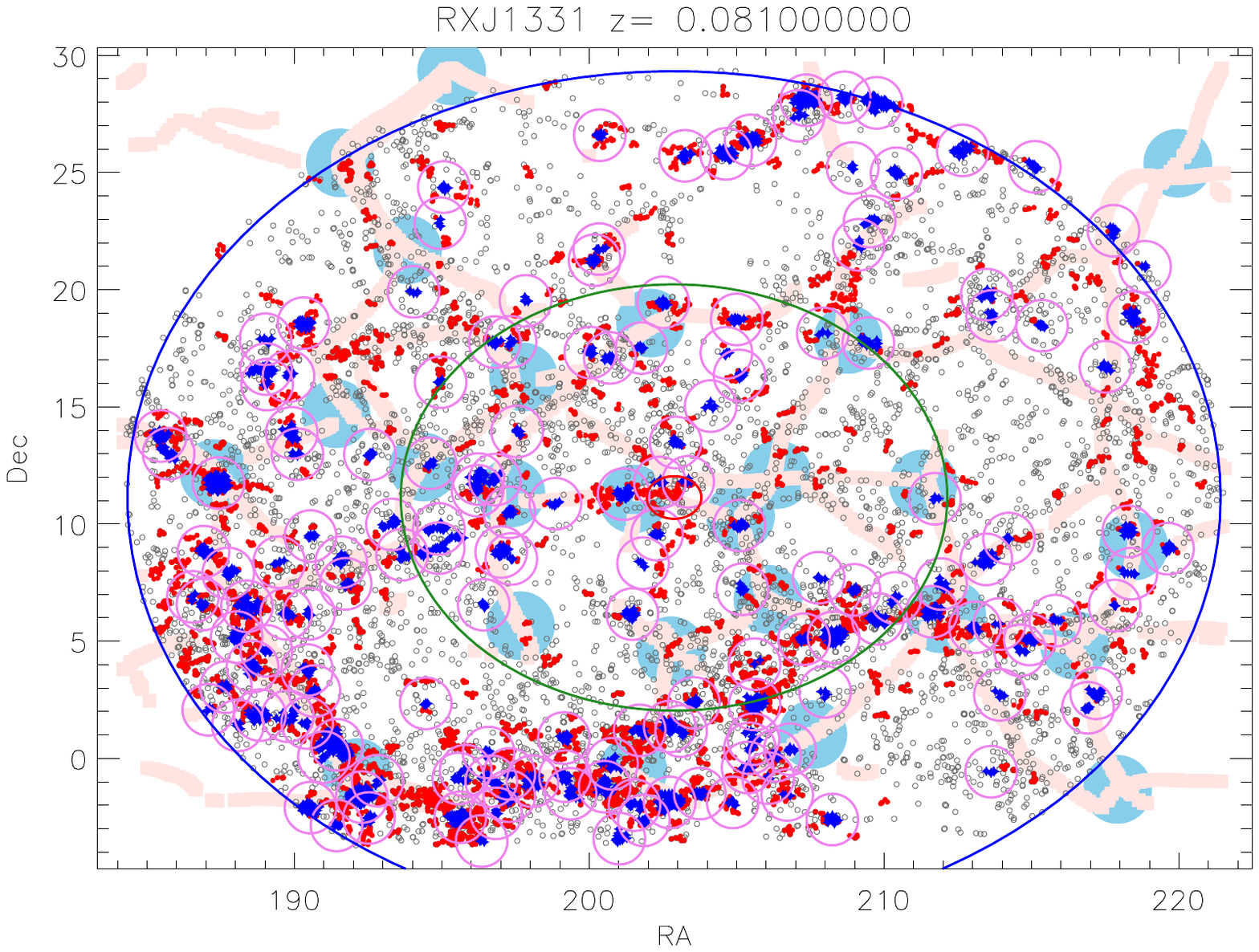}
  \end{tabular}
  
\end{figure*}

  \begin{figure*}
    \centering
\continuedfloat
\caption{Continued}
\begin{tabular}{@{}c@{}}
    \includegraphics[trim=0 450 0 80, width=0.99\textwidth]{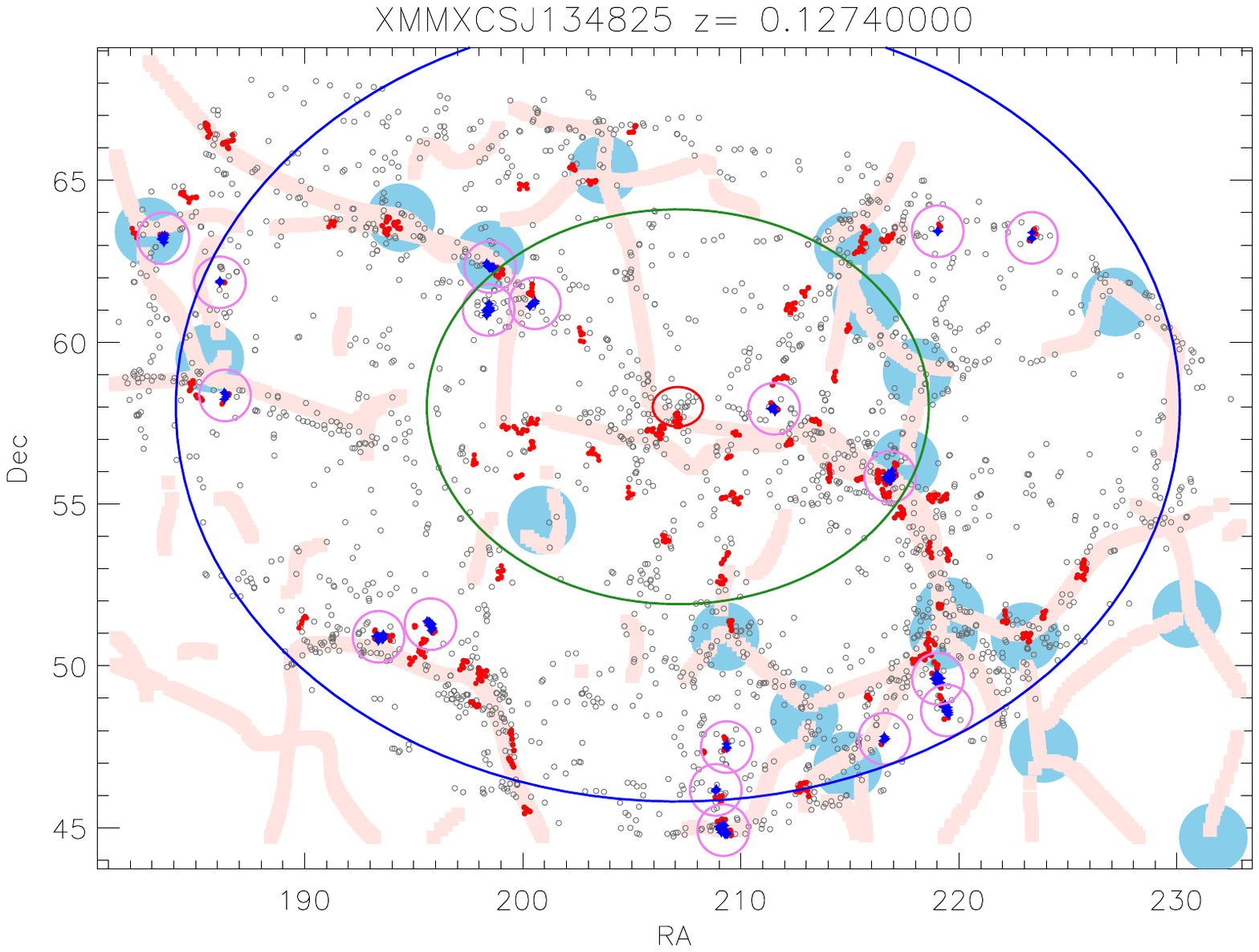}
  \end{tabular}

  \vspace{\floatsep}

  \begin{tabular}{@{}c@{}}
    \includegraphics[width=0.99\textwidth,trim=0 0 0 0]{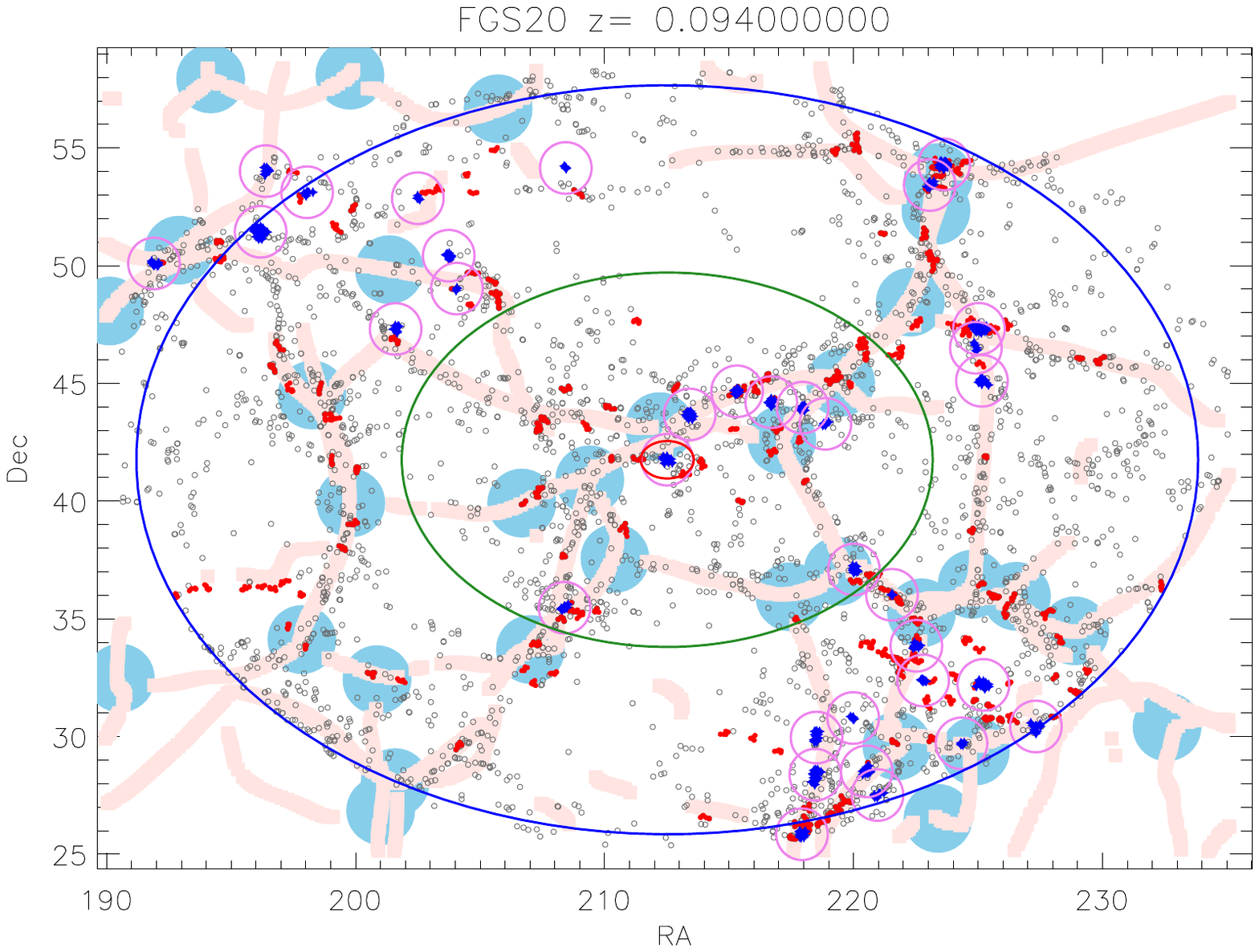}
  \end{tabular}
  
\end{figure*}

  \begin{figure*}
    \centering
\continuedfloat
\caption{Continued}
\begin{tabular}{@{}c@{}}
    \includegraphics[trim=0 450 0 80, width=0.99\textwidth]{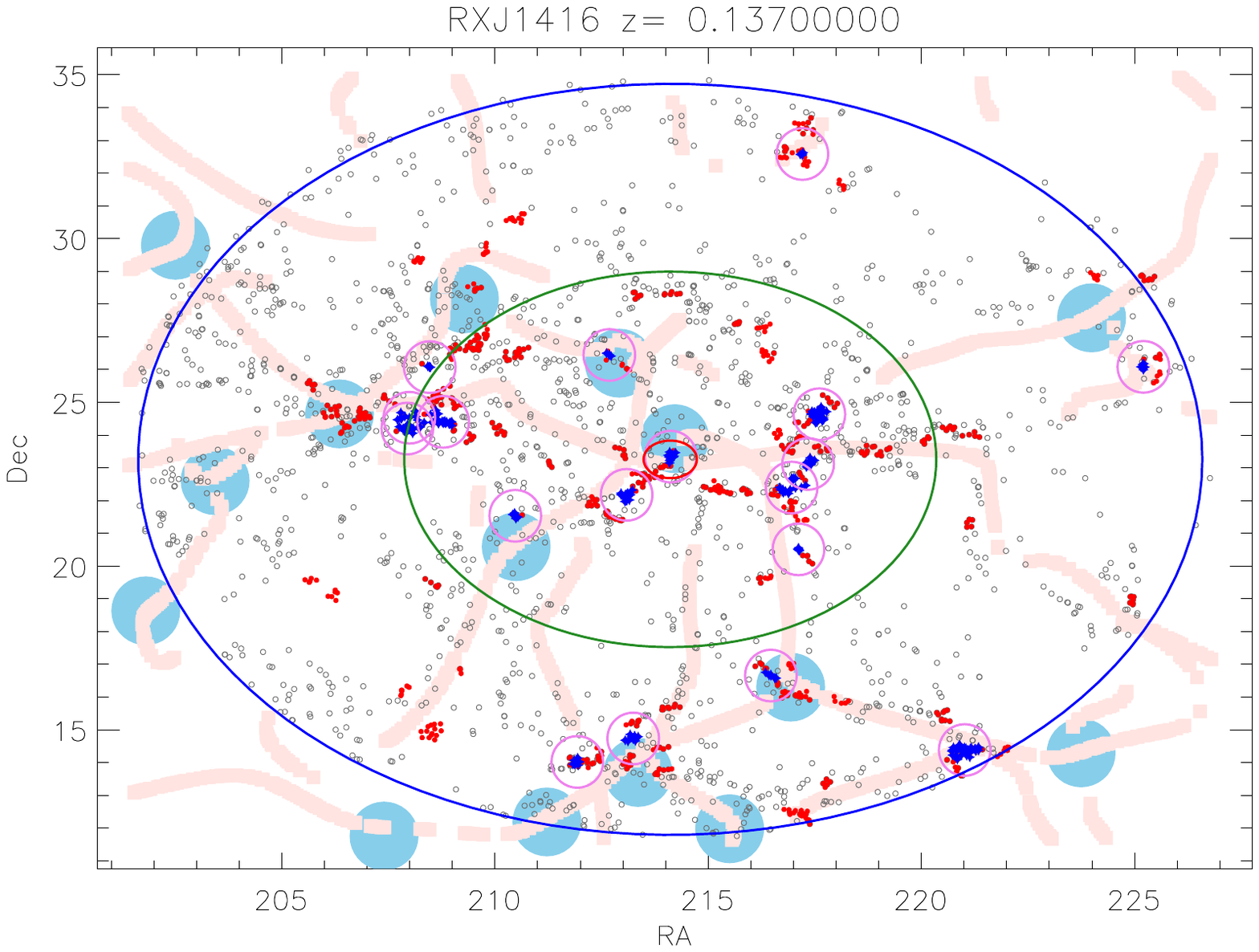}
  \end{tabular}

  \vspace{\floatsep}

  \begin{tabular}{@{}c@{}}
    \includegraphics[width=0.99\textwidth,trim=0 0 0 0]{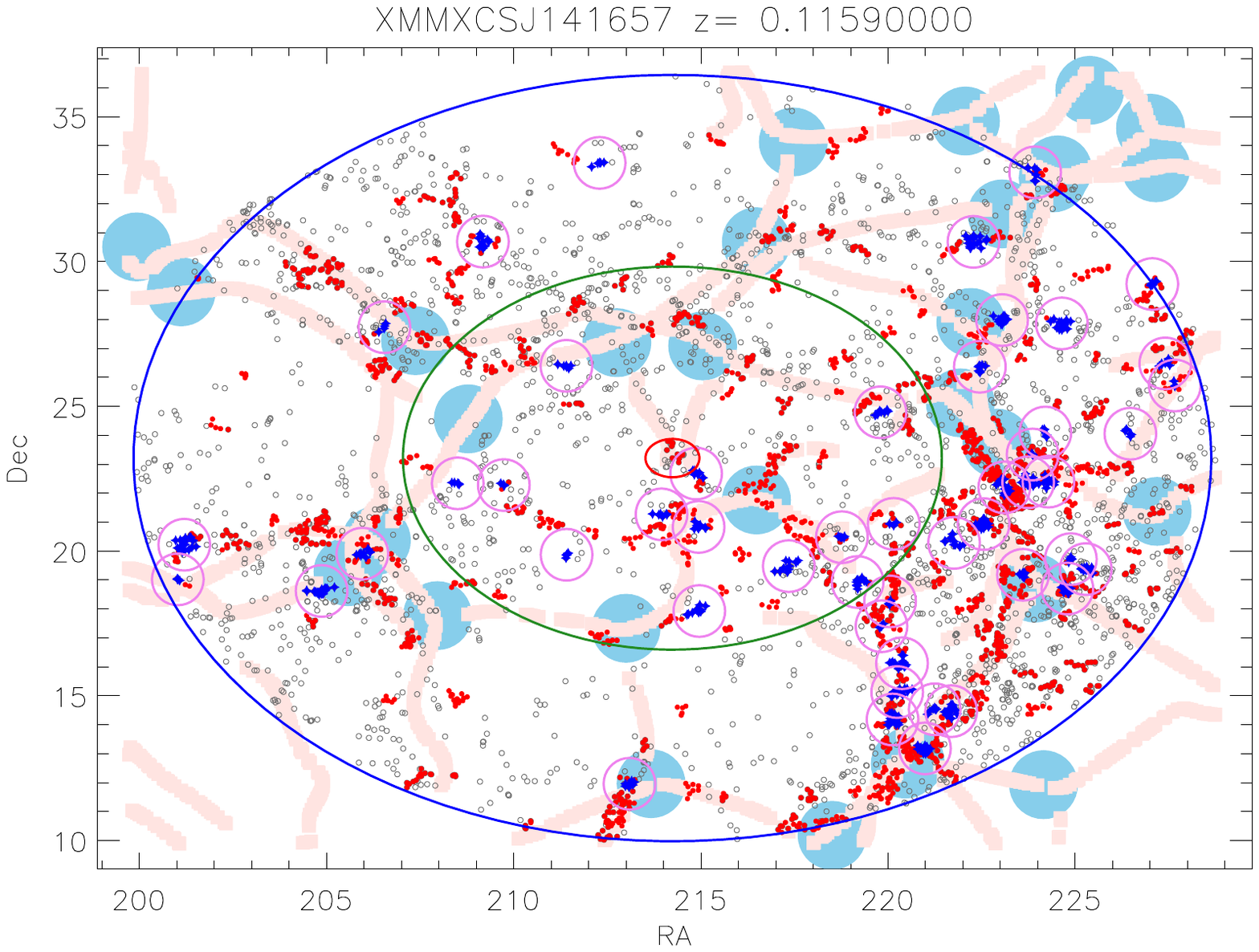}
  \end{tabular}
  
\end{figure*}

  \begin{figure*}
    \centering
\continuedfloat
\caption{Continued}
\begin{tabular}{@{}c@{}}
    \includegraphics[trim=0 450 0 80, width=0.99\textwidth]{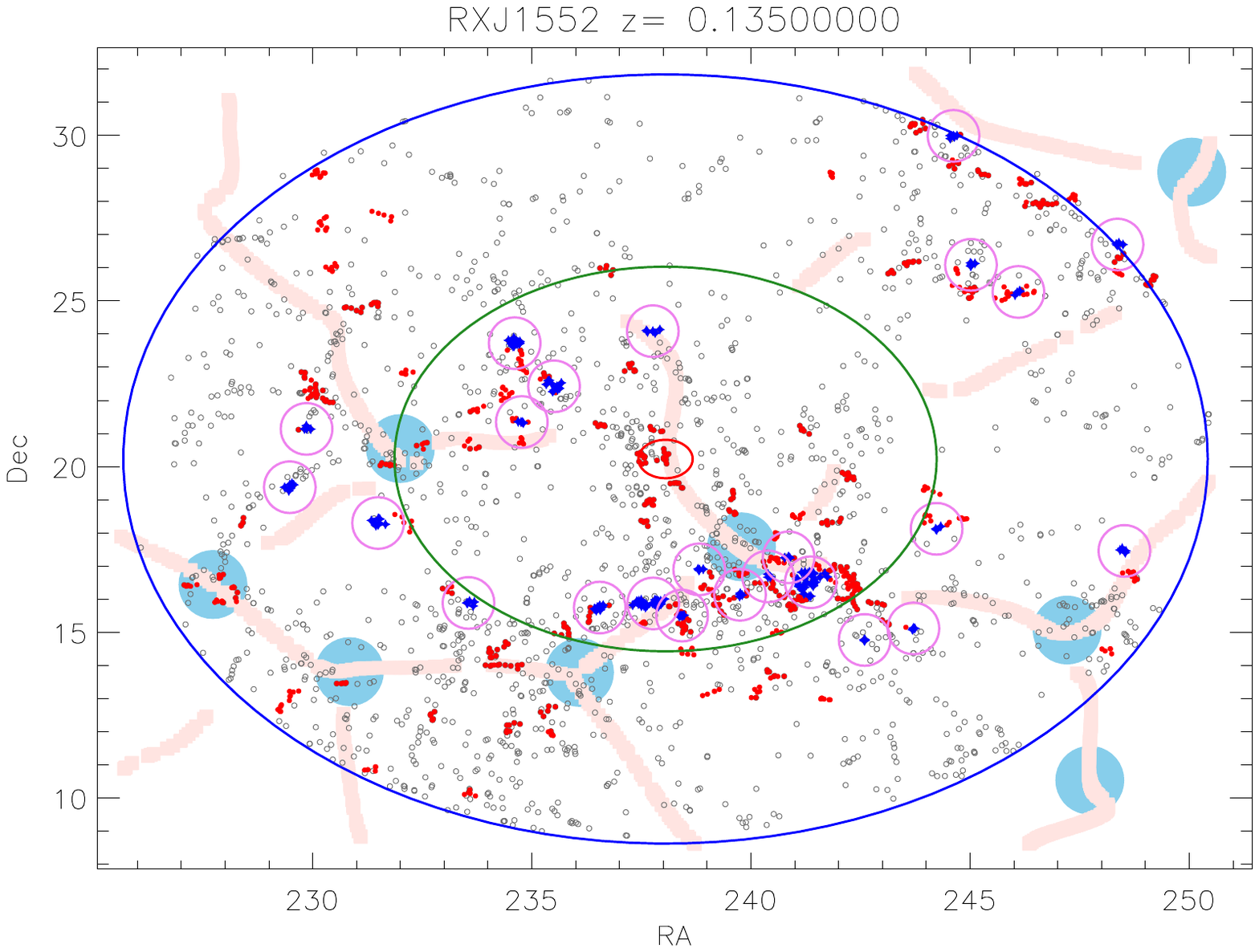}
  \end{tabular}

  \vspace{\floatsep}

  \begin{tabular}{@{}c@{}}
    \includegraphics[width=0.99\textwidth,trim=0 0 0 0]{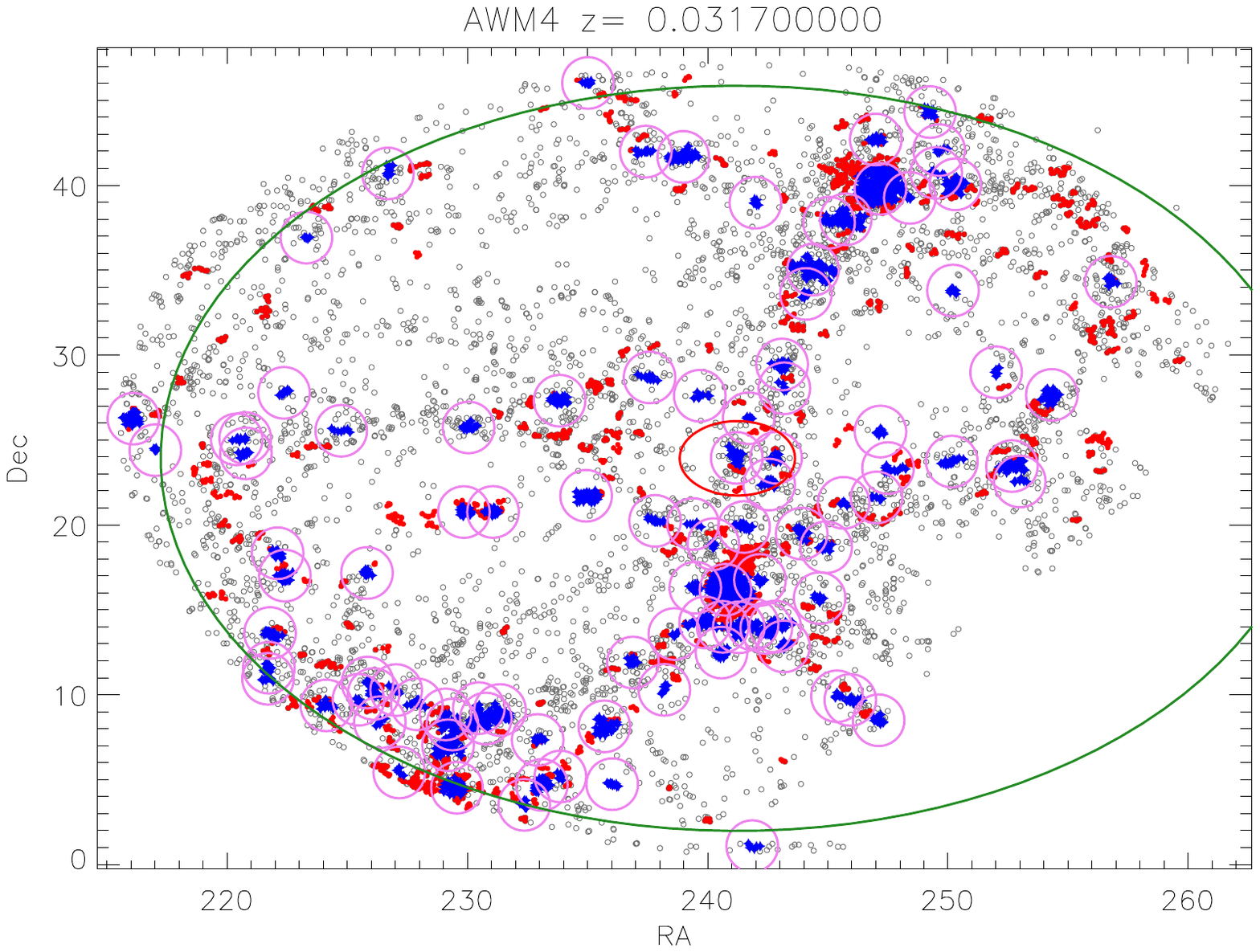}
  \end{tabular}
  
\end{figure*}

\newpage

\section{The large-scale structure of FGS28}
\label{appendixB}
In Fig. \ref{fig:FGS28} we present the large scale structure around FGS28. As we mentioned in Sect. \ref{sec:FGS28}, we did not include this FG into our sample for different reasons. However, since we mentioned the supercluster and filament that are found very close to FGS28 in both apparent position and redshift, we include the figure for sake of clarity.

\vspace{2cm}

\noindent\begin{minipage}{\textwidth}
    \captionof{figure}{The large scale structure around FGS28. Galaxies with SDSS spectroscopy within $\pm 1500$ km s$^{-1}$ from FGS28 redshift are represented in black. Moreover, in the bottom-left panel, the large red, green, and blue ellipses represent 5, 50, and 100 Mpc respectively (as in Fig. \ref{fig:lss-appendix}). The small violet, brown, green, and blue ellipses are the four clusters that are found close to FGS28. In the top-right panel, a zoom can be seen, were the color code is the same with the exception of the red ellipse, that now simply identify the position of FGS28. In this panel, the legend associate each coloured ellipse to a specific clusters. It can be seen that FGS28 is identified by a single galaxy with spectroscopic redshift in SDSS, supporting the idea that this is a isolated galaxy found in the surroundings of a large supercluster.}
    \centering
    \includegraphics[width=0.99\hsize]{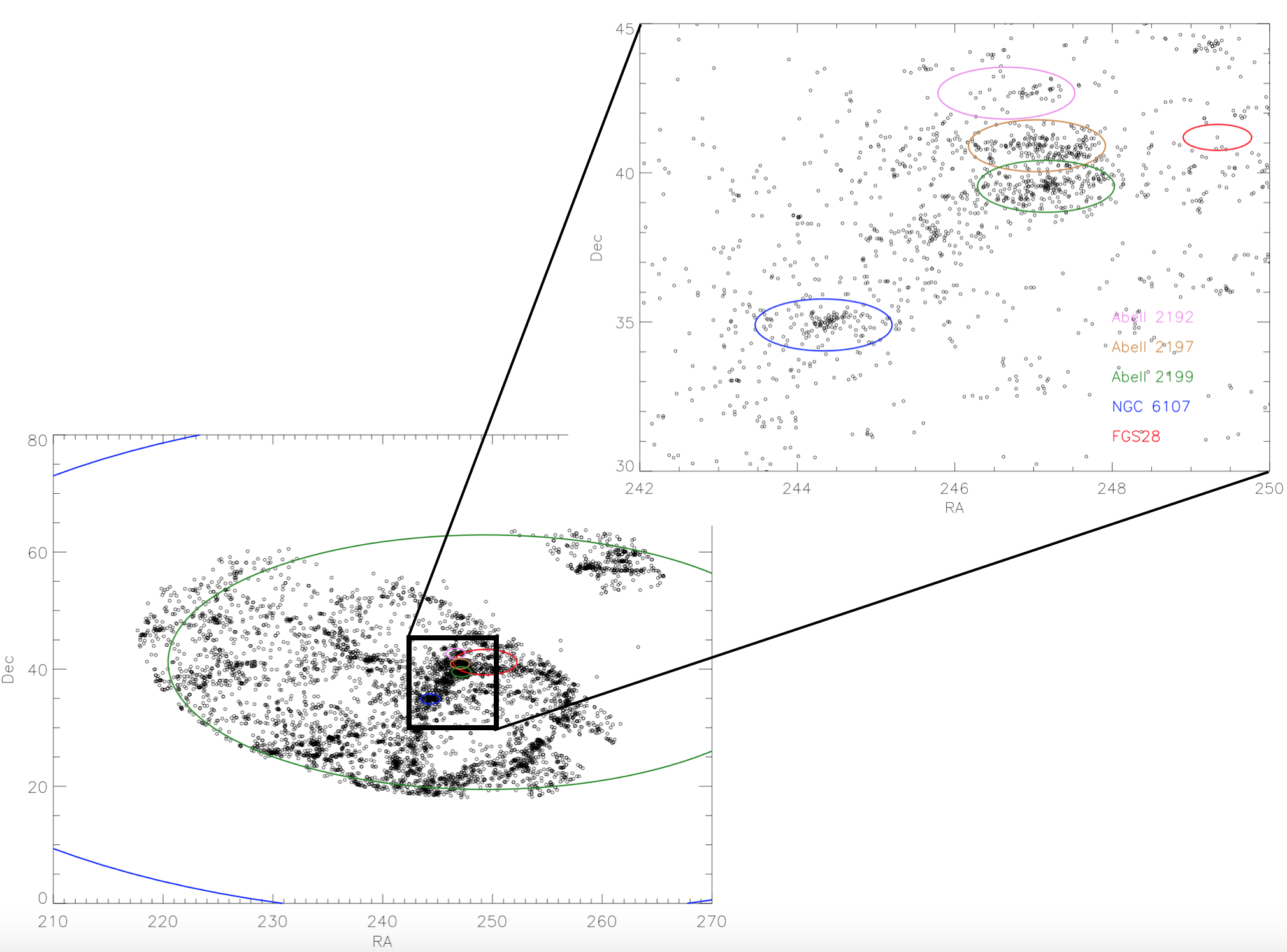}
    \label{fig:FGS28}
\end{minipage}

\end{document}